\documentclass[useAMS,usenatbib]{mn2e}

\usepackage{graphicx,amsmath,color}

\title[The infall of X-ray groups onto clusters]{LoCuSS: The infall of X-ray groups onto massive clusters}
\author[C. P. Haines et al.]{C.~P. Haines$^{1,2}$\thanks{E-mail:
chris.haines@brera.inaf.it}, A. Finoguenov$^{3,4}$, G.~P. Smith$^{5}$, 
A. Babul$^6$, E. Egami$^7$, 
\newauthor P. Mazzotta$^{8}$,  N. Okabe$^9$, M.~J. Pereira$^{7}$, M. Bianconi$^{5}$, S.~L. McGee$^{5}$,  F. Ziparo$^{5}$, 
\newauthor L. E. Campusano$^{2}$, C. Loyola$^{2}$\\
$^{1}$INAF - Osservatorio Astronomico di Brera, via Brera 28, 20121 Milano, Italy\\
$^{2}$Departamento de Astronom\'{i}a, Universidad de Chile, Casilla
36-D, Correo Central, Santiago, Chile\\
$^{3}$Max-Planck-Institut f\"{u}r extraterrestrische Physik, Giessenbachstraße, 85748 Garching, Germany\\
$^{4}$Department of Physics, University of Helsinki, Gustaf 
H\"{a}llstr\"{o}min katu 2a, FI-0014 Helsinki, Finland\\
$^{5}$School of Physics and Astronomy, University of Birmingham,
Edgbaston, Birmingham, B15, 2TT, UK\\
$^{6}$Department of Physics and Astronomy, University of Victoria,
3800 Finnerty Road, Victoria, BC, V8P 1A1, Canada\\
$^{7}$Steward Observatory, University of Arizona, 933 North Cherry
Avenue, Tucson, AZ 85721, USA\\
$^{8}$Dipartimento di Fisica, Universit\`{a} degli Studi di Roma 'Tor
Vergata', via della Ricerca Scientifica 1, 00133 Roma, Italy\\
$^{9}$Department of Physical Science, Hiroshima University, 1-3-1
Kagamiyama, Higashi-Hiroshima, Hiroshima 739-8526, Japan
}
\begin{document}


\pagerange{\pageref{firstpage}--\pageref{lastpage}} \pubyear{2017}

\maketitle

\label{firstpage}

\begin{abstract}
Galaxy clusters are expected to form hierarchically in a $\Lambda$CDM universe, growing primarily through
mergers with lower mass clusters and the continual accretion of
group-mass halos. Galaxy clusters assemble late, doubling their masses
since $z{\sim}0.5$, and so the outer regions of clusters should be
replete with infalling group-mass systems.  
We present an {\em XMM-Newton} survey to search for X-ray groups in
the infall regions of 23 massive galaxy clusters 
($\langle{\rm M}_{200}\rangle{\sim}10^{15\,}{\rm M}_{\odot}$) at $z{\sim}0.2$,
identifying 39 X-ray groups that have been spectroscopically confirmed to lie
at the cluster redshift. These groups have mass estimates in the range $2{\times}10^{13} -
7{\times}10^{14}\,{\rm M}_{\sun}$, and group-to-cluster mass
ratios as low as 0.02. The comoving number density of X-ray groups in
the infall regions is ${\sim}25{\times}$ higher than that seen for
isolated X-ray groups from the XXL survey. 
The average mass per cluster contained within these X-ray groups is
$2.2{\times}10^{14\,}{\rm M}_{\odot}$, or $19{\pm}5$\% of the mass
within the primary cluster itself. We estimate that
${\sim}10^{15\,}{\rm M}_{\odot}$ clusters increase their masses by $16{\pm}4$\%
between $z{=}0.223$ and the present day due to the accretion of groups with
$M_{200}{\ge}10^{13.2\,}{\rm M}_{\odot}$. This represents about half
of the expected mass growth rate of clusters at these late epochs. The
other half is likely to come from smooth accretion of matter not
bound within halos.
The mass function of the infalling X-ray groups appears significantly
top heavy with respect to that of ``field'' X-ray systems, consistent
with expectations from numerical simulations, and the basic
consequences of collapsed massive dark matter halos being biased
tracers of the underlying large-scale density distribution. 

\end{abstract}


\section{Introduction}

A key prediction of $\Lambda$CDM cosmological models is that structure
formation occurs hierarchically, whereby dark matter (DM) halos grow via
the continual accretion of lower mass systems. More massive halos thus
on average form later than less massive halos. Galaxy clusters as
the most massive collapsed halos form latest, doubling their masses on
average since $z{\sim}0.5$ \citep{boylan,gao}, and are also the most dynamically
immature. This manifests itself in 40\% of local clusters showing clear 
substructure in their X-ray emission \citep{jones,chon12,mann}, and a
higher fraction with substructure in their underlying mass
distributions inferred from gravitational lensing \citep{smith05,martinet}.

The abundances and rates of growth of galaxy
clusters thus represent sensitive probes of cosmology \citep{voit,henry,kravtsov}, providing
constraints on the primary cosmological parameters
($\Omega_{m}$, $\Omega_{\Lambda}$, $\sigma_{8}$, $w_{0}$) that are
competitive with, and complementary to, those from supernovae, CMB and
baryon acoustic oscillations \citep{allen}. While measurements of the
local abundance and mass function of clusters have been used to
jointly constrain the cosmic matter density $\Omega_{m}$ and the
amplitude of density perturbations $\sigma_{8}$ \citep[e.g.][]{schuecker,planckXX}, the rate of growth of
massive clusters provides a window on the form and content of dark energy via
the impact of cosmic acceleration on both structure formation and the
distance-redshift relation \citep{mantz10}. 
\citet{vik09} compared the cluster mass function at $z{\sim}0.5$ with
that of present day clusters, finding an average mass growth of
75--80\% for massive clusters between $z{\sim}0.5$ and 0.05. This
evolution maps the growth of structure and provides strong constraints
on the dark energy density of the Universe with
$\Omega_{\Lambda}{=}0.83{\pm}0.15$ (i.e. non-zero at ${>}5\sigma$ significance) 
 and its equation-of-state parameter $w_{0}{=}{-}1.14{\pm}0.21$, assuming
 a constant $w$ and a flat universe.

The merger histories of dark matter halos and the mass
functions of the progenitor halos that are accreted by the primary
cluster halo have been derived analytically using the extended
Press-Schechter formalism \citep{bond,bower,lacey} and through cosmological
N-body simulations \citep*{lemson,governato,giocoli08} with good agreement
between the two approaches. 
Interestingly, the {\em unevolved} subhalo mass function
d$N(m_{sub}/M_{0})/{\rm d}\,\ln(m_{sub}/M_{0})$, which parametrizes the masses of
the progenitor subhalos ($m_{sub}$) at the time they were accreted onto the primary cluster halo, has been
found to be universal, with no dependency on the primary halo mass
$M_{0}$, redshift \citep{van05,giocoli08} or even the cosmological
parameters \citep{zentner,yang}. 

Collapsed dark matter halos are biased tracers of the underlying
matter distribution, with the most massive halos forming from the
highest peaks in the primoridal linear density field. The extended
Press-Schechter formalism provides robust predictions for this bias as
an increasing function of peak height $\nu$ or equivalently halo mass \citep{mo,tinker10}.
This leads to a systematic variation of halo mass function with
large-scale (${\sim}10$\,Mpc) environment, with high-mass halos over-represented in high-density regions
\citep{mo}, a prediction which is reproduced well in N-body
simulations \citep{lemson,governato,faltenbacher}. As clusters are preferentially located at the centres of
large-scale overdensities, the mass function of halos in their
surroundings, and which will subsequently be accreted onto them, should be
biased towards higher mass systems. 

The late assembly of clusters implies that there must be numerous
group-mass systems in their outskirts in the process of being accreted. 
This motivates an X-ray survey covering the infall regions of a
representative sample of massive clusters aimed at detecting
these infalling group-mass systems and estimating their masses, with
the ultimate objectives of estimating how much mass they contribute to
the growth of the cluster through accretion, and whether their mass
function is indeed signficantly biased with respect to more typical
regions of the Universe. A key advantage of detecting these groups
from their extended X-ray emission is that they can be unambiguously
identified as massive virialized DM halos. 
In contrast, a purely optically-selected group sample \citep{ragone,lemze} could be
biased by the inclusion of non-virialized systems or chance
line-of-sight projections \citep{pearson,osullivan}, the likelihood of which are dramatically
increased in the vicinity of rich clusters.

Many massive clusters at $z{\sim}0.2$ have been observed by {\em XMM-Newton}
with exposure times ${\ga}10$\,ksec, sufficient to obtain reliable
temperature, gas density and mass profiles out to $r_{500}$ and
M$_{500}$ mass estimates accurate to 10--25\% \citep{zhang07,martino}. These depths also
permit the detection of galaxy groups at $z{\sim}0.2$ down to masses of
${\sim}$2--3${\times}10^{13}{\rm M}_{\odot}$. 
While the ICM of the primary cluster is usually only detectable out to
${\sim}r_{500}$, the 30\,arcmin field-of-view of {\em XMM-Newton}
provides coverage out to ${\sim}1$.5--2.0\,$r_{200}$ and enabling infalling
groups in the cluster outskirts to be detected. 

In this article we present a search for X-ray groups in
existing {\em XMM-Newton} X-ray observations targetting 23 massive clusters at
$z{\sim}0.2$ from the Local Cluster Substructure Survey (LoCuSS), for
which highly-complete stellar mass limited optical spectroscopy is
available from the Arizona Cluster Redshift Survey \citep[ACReS;][Pereira et al. in preparation]{haines13,haines15}.

With this {\em XMM} group sample, we derive the mass
function of galaxy groups infalling into massive clusters down to
M$_{200}{\sim}2{\times}10^{13}\,{\rm M}_{\odot}$, and
estimate whether the accretion of these groups onto the clusters is
sufficient to explain the expected mass growth of the clusters between
$z{\sim}0.2$ and the present day, or if further sources such
as smooth accretion of dark matter are required. We examine the
group-cluster mass ratio distribution and compare it to the unevolved
subhalo mass function, whose universality is a key prediction of $\Lambda$CDM
cosmological models \citep[e.g.][]{giocoli08}. 

We use a $\Lambda$CDM cosmology with $\Omega_{M}{=}0.27$,
$\Omega_{\Lambda}{=}0.73$ and $H_{0}{=}72\,h_{72}$\,km\,s$^{-1}$\,Mpc$^{-1}$. 

\section{The cluster sample, XMM data and construction of the group catalogue}

\subsection{The primary cluster sample}
 
The primary cluster sample for this study consists of all 23 X-ray luminous clusters 
 within LoCuSS for which there is {\em both} high-quality {\em
   XMM-Newton} X-ray data {\em and} extensive spectrosopic coverage of cluster
 galaxies out to ${\sim}3\,r_{200}$ from ACReS to identify the most luminous member galaxies within the group and
 securely confirm its redshift.

LoCuSS is a systematic multi-wavelength survey of ${\sim}100$ X-ray selected ($L_{X}{\ge}\,2{\times}10^{44\,}{\rm erg\,s}^{-1}$) massive clusters
at $0.15{\le}z{<}0.30$, drawn from the {\em ROSAT} All Sky Survey
catalogues \citep[RASS;][]{ebeling98,ebeling00,bohringer}. The ACReS
subsample consists of the first batch of 30 clusters having wide-field
optical imaging out to the virial radii from Subaru/Suprime-Cam and 
{\em Hubble Space Telescope} imaging of the cluster cores, enabling
detailed mass maps combining weak- (Subaru) and strong-lensing (HST)
data. All 30 systems have excellent ancillary wide-field,
multi-wavelength data including near-infrared imaging with
UKIRT/WFCAM ($J,K$; $52^{\prime}{\times}52^{\prime}$), {\em Spitzer}/MIPS
24$\mu$m and {\em Herschel}/PACS+SPIRE far-infrared photometry over
$25^{\prime}{\times}25^{\prime}$ fields \citep{haines10,smith10}. 
The $L_{X}$ distribution of the ACReS subsample is statistically indistinguishable
from the parent volume-limited sample of {\em ROSAT} clusters \citep{okabe}.
Of the 30 clusters covered by ACReS, 23 have existing {\em XMM-Newton} X-ray data
suitable for detecting group-mass systems at the redshift of the cluster.
These 23 systems form our primary cluster sample and are listed in
Table~\ref{clusters}.

\begin{table}
\caption{The primary cluster sample. Cols. (1,2) Cluster name and
  redshift. Col. (3) {\em ROSAT} 0.1--2.4\,keV
  X-ray luminosity, except A689$^{a}$ which comes from \citet{giles}.  
 Col. (4) $M_{200}$ masses from the {\em Chandra-XMM} analysis of \citet{martino}
for the high-$L_{X}$ cluster sample, extending the mass profiles out
to $r_{200}$, except for A665 and A2218$^{c}$ which come from
\citet{haines13}. 
Col. (5) Weak lensing $M_{200}$ estimates from \citet{okabe15}, except
for A665, A689 and A2218$^{d}$ whose mass estimates come from
\citet{pedersen, okabe, mahdavi}.} 
\centering
\begin{tabular}{lccr@{$\,\pm\,$}lr@{$\,\pm\,$}l} \hline
Cluster & z & $L_{X}$ & \multicolumn{2}{c}{$M_{200,X}$} & \multicolumn{2}{c}{$M_{200,WL}$} \\
Name & &\!\!\!\!($10^{44}$\,erg/s)\!\!\!\!& \multicolumn{2}{c}{($10^{14}\,{\rm M}_{\odot}$)} & \multicolumn{2}{c}{($10^{14}\,{\rm M}_{\odot}$)} \\ \hline
Abell 68     & 0.251 & 9.473 & 17.09 & 6.45   &   9.24 & 1.74\\
Abell 115N\! & 0.192 & 8.895 &   6.24 & 1.18$^b$\!& 9.78 & 3.22\\
Abell 209   & 0.209 & 6.289 &   8.71 & 1.74   & 17.71 & 2.90\\
Abell 267   & 0.229 & 8.569 &   9.28 & 2.83   &   8.29 & 1.56\\
Abell 291   & 0.196 & 4.883 &   4.16 & 0.73   &   7.82 & 2.01\\
Abell 383   & 0.189 & 4.559 &   4.46 & 1.20   &   7.26 & 1.65\\
Abell 611   & 0.286 & 8.855 & 10.66 & 1.67   & 12.82 & 2.52\\
Abell 665   & 0.183 & 9.837 & 11.71 & 2.79$^c$\!& 12.11 & 6.83$^d$\\
Abell 689   & 0.278 & \,\,\,1.812$^{a}$ &4.34  & 0.45 &  1.82 & 0.90$^d$\\
Abell 697   & 0.282 & 10.57 & 21.28 & 4.34   & 13.53 & 3.49\\
Abell 963   & 0.204 & 6.390 &   8.92 & 2.01   &   9.90 & 1.79\\
Abell 1689 & 0.185 & 14.07 & 14.93 & 3.69   & 15.25 & 2.17\\
Abell 1758N\!\!\!\!&0.279 & 7.514 & 18.21 & 3.59   &   8.17 & 1.98\\
Abell 1763 & 0.232 & 9.317 & 14.50 & 3.18   & 23.50 & 4.25\\
Abell 1835 & 0.252 & 24.48 & 21.46 & 5.68   & 14.01 & 2.43\\
Abell 1914 & 0.167 & 10.98 & 10.70 & 2.41   & 12.12 & 2.44\\
Abell 2218 & 0.173 & 5.554 &   6.73 & 1.60$^c$\!&   7.49 & 1.99$^d$\\
Abell 2219 & 0.226 & 12.73 & 21.22 & 4.94   & 14.01 & 2.71\\
Abell 2390 & 0.229 & 13.43 & 22.13 & 7.11   & 14.72 & 2.49\\
RXJ\,1720  & 0.160 & 9.573 &   9.12 & 3.81   &   7.26 & 2.37\\
RXJ\,2129  & 0.234 & 11.66 &   7.19 & 0.86   &   6.51 & 2.03\\
ZwCl 2089 & 0.235 & 6.786 &   2.67 & 0.35   &   3.40 & 1.39\\
ZwCl 7160 & 0.257 & 8.411 &   6.02 & 1.37   &   6.14 & 2.84\\ \hline
\end{tabular} 
\label{clusters}
\end{table}
 
Twenty out of the 23 clusters form part of the
volume-limited ``high-$L_{X}$'' LoCuSS sub-sample of 50 systems with 
\mbox{$L_{X}(0$.1--2.4\,keV$)/E(z){\ge}\,4.2{\times}10^{44\,}$erg\,s$^{-1}$, 
${-}25^{\circ}{<}\delta{<}{+}65^{\circ}$} and 
$n_{H}{\le}7{\times}10^{20}\,{\rm cm}^{-2}$ of \citet{okabe13,okabe15}
and \citet{martino}.
\citet{martino} extracted gas density and de-projected temperature
profiles for each ``high-$L_{X}$'' cluster from the same {\em XMM} images analysed
here, and derived total gravitational mass profiles and $M_{500}$
masses assuming hydrostatic equilibrium. These mass profiles were extended out to
$r_{200}$ producing the M$_{200,X}$ mass estimates used here
(Col. 4), and have mean mass uncertainties $\langle
dM_{200}/M_{200}\rangle{=}0.238$ (0.093\,dex). These uncertainties are
almost double those reported by \citet{martino} for the $M_{500}$
masses (0.052\,dex), due to the need to extrapolate the mass profiles beyond
$r_{500}$, but the ratios between these two masses vary little between
systems, $(M_{500}/M_{200}){=}0.648{\pm}0.094$. These 20 clusters also
have updated weak-lensing mass measurements from \citet{okabe15}
(Col. 5) that are fully consistent on average with our X-ray mass
estimates \citep{smith16}, with geometric mean mass ratio 
$\langle M_{200,X}/M_{200,WL}\rangle{=}0.985{\pm}0.106$.

Of the remaining three clusters, two (A665 and A2218) were only excluded from the
high-$L_{X}$ sample due to their declination ($\delta{\sim}{+}66^{\circ}$), and 
we take their X-ray mass estimates (M$_{200,X}$) from
\citet{haines13}. These are based on fitting the phenomenological cluster
models of \citet{ascasibar} to a series of annular spectra extracted
from deep {\em Chandra} data for each cluster \citep{sanderson10}.
The {\em ROSAT} $L_{X}$ estimate of the final cluster Abell\,689 satisfied
our high-$L_{X}$ selection and it was included in the ACReS sample of
30 clusters. It was excluded from the ``high-$L_{X}$'' sample
of \citet{martino} however, as the {\em Chandra} analysis of \citet{giles}
showed that its X-ray emission is dominated by a central BL Lac, and
after excluding this central point source, the $L_{X}$ from the
extended cluster emission falls below the LoCuSS survey limit. 
ACReS does confirm A689 as a cluster \citep[with 338
members;][]{haines15}, and we estimate its
mass from its updated $L_{X}$ using the same $M_{200}{-}L_{X}$ relation \citep{leauthaud} as used later for our
{\em XMM}-detected groups ({\S}\ref{sec:catalogue}; Eq.~\ref{m200_lx}).

A115 is a complex cluster merger with two
approximately equal mass components A115N and A115S \citep{okabe,gutierrez}. A115N is more
X-ray luminous than A115S \citep{forman}, and so we take A115N as our primary
cluster, while
\citet{martino} only measured the mass of A115S. For consistency, we
derive the $M_{200,X}$ values for both A115N and A115S from the $L_{X}$
estimates obtained in our analysis of the {\em XMM} data
($\S$~\ref{sec:reduction}), using Eq.~\ref{m200_lx} ({\S}~\ref{sec:catalogue}).


\subsection{Detecting extended sources in the XMM data}
\label{sec:reduction}

The details of the {\em XMM-Newton} observations and initial data reduction are 
summarized in \citet{martino}.
After removal of energies affected by instrumental lines \citep[as in][]{finoguenov07},
the 0.5--2\,keV band images from the pn and MOS detectors are
in-field background subtracted and co-added. 
 To detect and identify extended emission from X-ray groups, the 
0.5--2\,keV band image for each cluster is then decomposed into unresolved
and extended sources, using the wavelet scale-wise decomposition and
reconstruction technique of \citet{vik98}, employing angular scales
from 8--64$^{\prime\prime}$. Similarly to
\citet{finoguenov09,finoguenov10,finoguenov15} point sources in the
{\em XMM} images are detected using the scales of 8$^{\prime\prime}$
and 16$^{\prime\prime}$ down to a wavelet significance of 4$\sigma$.
The full flux of each detected point source is reconstructed using the {\em XMM}
PSF model and removed from the image, following \citet{finoguenov09}. 

Having removed point sources, we apply the extended source search
algorithm, applying the wavelet detection at 32$^{\prime\prime}$ and
64$^{\prime\prime}$ scales, and generating a noise map corresponding to the
32$^{\prime\prime}$ scale against which the
extended flux in the reconstructed image is tested for significance.
We consider extended sources of X-ray emission as those detected above a $4\sigma$
threshold in the wavelet analysis, relative to the level of background
fluctuations. The primary target of the observations (LoCuSS clusters) 
has been automatically detected as a part of this procedure. 
The flux measurements of each detected source are made within
elliptical apertures, forcing SExtractor to 
describe the wavelet image as it is. The extent of the emission is 
traced by the wavelet routine down to 1.6$\sigma$. At the depths of these {\em XMM} observations, the
actual signal-to-noise ratio (SNR) of the detected sources extend down
to ${\sim}2\sigma$, as both the contribution of the source flux to the
noise is non-negligible and the flux extraction extends to areas where
the source significance is just 1.6$\sigma$ over the background level.


\subsection{Determination of X-ray group redshifts}

The optical counterparts of extended X-ray sources were identified using the
combination of deep optical imaging and extensive spectroscopic data
(ACReS), for which we have complete coverage over all 23 {\em XMM} fields. 
Each cluster was observed with Suprime-Cam
($34^{\prime}{\times}27^{\prime}$ field-of-view) on the 8.2m Subaru
telescope to perform the weak lensing analysis of \citet{okabe,okabe13,okabe15}.
Typical observations consisted of two bands ($V,i$) with 30--40\,min exposure times and FWHM${\sim}0.7^{\prime\prime}$, 
providing high-quality photometry down to $i_{AB}{\sim}26$. 
The Arizona Cluster Redshift Survey
\citep[ACReS;][]{haines13,haines15} observed all 23 clusters in our
sample with Hectospec, a 300-fiber multi-object spectrograph with a 1-degree
diameter circular field of view that is installed on the 6.5m MMT
telescope. Target galaxies were primarily $K$-band selected down to a limit
of $m_{K}^{*}(z_{cl}){+}2.0$ to produce an approximately stellar
mass-limited sample down to $\mathcal{M}{\sim}1.5{\times}10^{10\,}{\rm
  M}_{\odot}$, with a $J{-}K$ colour selection used to efficiently target
galaxies at approximately the redshift of the primary cluster
\citep{haines09,haines09b,haines13}, irrespective of their
star-formation history (i.e. with no bias towards red sequence or
star-forming galaxies).  
We achieve spectroscopic completeness levels of ${\sim}80$\%
for $M_{K}{<}{-}23.10$ ($M_{K}^{*}{+}1.5$) cluster galaxies within the
23 {\em XMM} fields. 

For each X-ray source, contours of the extended X-ray emission
are overlaid on the Subaru optical images.
In many cases there is a clear dominant early-type galaxy located near
the centre of the X-ray emission, which we take to be the central
group galaxy. We have a spectroscopic redshift of this galaxy for all the candidate $z{\la}0.4$ groups. 
Further group members are then sought as fainter
galaxies with redshifts within 1000\,km\,s$^{-1}$ of the central
galaxy, located within 2--3\,arcmin. 
In cases where no dominant central galaxy is visible, we seek at least
two galaxies within 1--2\,arcmin of the X-ray centre with redshifts within
1000\,km\,s$^{-1}$ of each other. The group redshift is taken to be
the mean redshift of its member galaxies. More details of the process
to determine group members is given in \citet{bianconi}.

\begin{figure}
\centerline{\includegraphics[width=84mm]{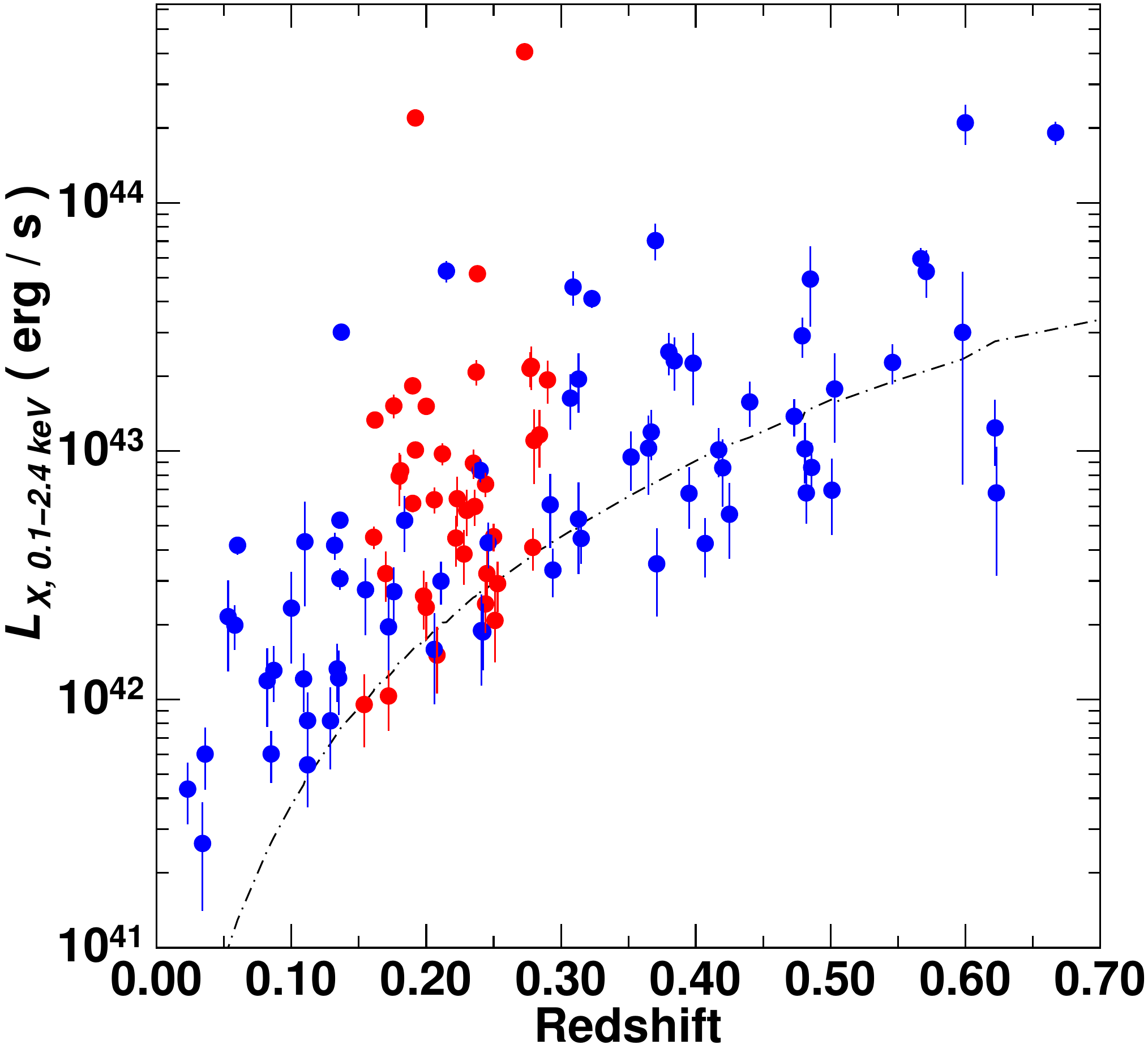}}
\caption{X-ray luminosity-redshift sampling of all the galaxy groups detected
  in the {\em XMM} images with a confirmed spectroscopic
  redshift. Those galaxy groups at the redshift of the primary
  cluster in the same {\em XMM} image are indicated in red, while blue
  points mark the remaining ``isolated'' galaxy groups. The dot-dashed
  curve indicates a flux threshold of $10^{-14}\,{\rm
    erg\,s}^{-1}\,{\rm cm}^{-2}$. }
\label{Lx_z}
\end{figure}

We are able to identify every single extended X-ray source detected
at ${\rm SNR}{>}6.0$ as a galaxy group with known redshift, comprising
32 X-ray groups at redshifts 0.06--0.67 in addition to the targetted
primary clusters.
At lower significance levels (mostly at ${<}3{\sigma}$), we find a number of the ``extended'' X-ray sources to
be centred on background QSOs, some of which had been previously 
identified as X-ray point sources in {\em Chandra} imaging \citep{haines12}.
Most of these QSOs had spectra having been observed as part of
ACReS strategy to target all mid-infrared bright sources
($f_{24}{>}1$\,mJy), including those unresolved in our $K$-band
imaging, resulting in a 24$\mu$m-selected sample of QSOs
\citep{xu,xu2}. 

\begin{table*}
\caption{The {\em XMM} sample of infalling galaxy groups. The complete
  table of groups will appear in the final published version. 
Cols. (1,2)
  IAU name and short ID of X-ray group. Cols. (3,4) Coordinates of
  the centroid of the group's X-ray emission ($\alpha, \delta$). Col. (5) Mean redshift of group
  members. Col. (6) Rest-frame X-ray luminosity of the group in the
  0.1--2.4\,keV band. Col. (7) Estimate of the cluster mass M$_{200}$
  in units of 10$^{13\,}{\rm M}_{\odot}$. Col. (8) Signal-to-noise
  ratio of the X-ray detection. Col. (9) Projected distance of group
  from the primary cluster centre in units of $r_{200}$. Col. (10)
  Flag indicating whether the
group contains a dominant central galaxy (BGG) or not. Col. (11) Number of
spectroscopically confirmed group members.}
\centering
\begin{tabular}{llcr@{:}l c r@{$\,\pm\,$}lr@{$\,\pm\,$}lrccr} \hline
& & Right & \multicolumn{2}{c}{} &
  & \multicolumn{2}{c}{$L_{X}$} &
\multicolumn{2}{c}{} & & & \\
IAU Name & Group & Ascension & \multicolumn{2}{c}{Declination} &
\multicolumn{1}{c}{$\langle z \rangle$}  & \multicolumn{2}{c}{(0.1--2.4\,keV)} &
\multicolumn{2}{c}{$M_{200}$} & SNR & \underline{$r_{proj}$} & \!\!\!BGG\!\!\!& $N_{z}$ \\
XMMU & ID & (J2000) & \multicolumn{2}{c}{(J2000)} & & \multicolumn{2}{c}{($10^{42\,}{\rm erg\, 
  s}^{-1}$)} & \multicolumn{2}{c}{($10^{13\,}{\rm M}_{\odot}$)} &  & $r_{200}$ &\!\!Y/N\!\!&\\ \hline
J013137.0-134501 & A209-g10 & 01:31:37.04 & --13:45 & 01.9 & 0.20034 & 2.35 & 0.62 &
3.03 & 0.68 & 3.8 & 1.020 & Y & 2\vspace{1mm}\\
J101640.1+385443 & A963-g10 & 10:16:40.13 & +38:54 & 43.0 & 0.20129 & 15.14 & 1.12
& 10.00 & 1.52 & 13.5 & 1.012 & Y & 21\vspace{1mm}\\
J133210.6+503031 & A1758-g7 & 13:32:10.67 & +50:30 & 31.5 & 0.27903 & 4.09 & 0.80 &
4.06 & 0.77 & 5.1 & 0.643 & Y & 18\\
J215309.4+174224 & A2390-g1 & 21:53:09.47 & +17:42 & 24.9 & 0.22184 & 4.46 & 1.04 &
4.49 & 0.93 & 4.3 & 0.584 & Y & 1\\ \hline

\end{tabular}
\label{grouplist}
\end{table*}

As the significance of the extended X-ray source declines, the
fraction for which we find no likely counterpart in the optical
images starts to increase until for 2.5${<}$SNR${<}$3.0 ${\sim}4$0\% of X-ray
sources remain unidentified. The bulk of these unidentified sources are
likely confused low-luminsity AGN \citep{finoguenov07} which are not
detected on the small (8--16$^{\prime\prime}$) scales. 
There are no galaxies that satisfy the ACReS target selection
criteria within 1\,arcmin of any of these unidentified X-ray sources. 
In other words, we do not expect to have missed identifying any $z{<}0.3$ groups due to incomplete
spectroscopy.   

Over the 23 {\em XMM} fields, excluding the primary clusters, we
identify a total of 91 X-ray groups above a SNR limit of 3.0,
with redshifts in the range 0.02--0.67. Ninety of these have
at least one member with a spectroscopic redshift. The remaining 
X-ray source is centred on a compact clump of red galaxies with $J-K$
colours consistent with the group being at $z{\sim}0.6$. 
Figure~\ref{Lx_z} shows the distribution of these 90 groups in the
$L_{X}$ versus redshift plane. X-ray groups that are at the same
redshift as the central cluster in their {\em XMM} field are indicated
in red, and can be typically detected down to $L_{X}(0$.1--2.4\,keV$){\sim}2{\times}10^{42\,}{\rm erg\,s}^{-1}$.


\subsection{Estimation of the X-ray group masses}
\label{sec:catalogue}

 Using the knowledge of the group's redshift, the global properties
($L_{X}$) of the 
groups are determined based on the detected flux within the aperture and 
estimating the total flux, based on the correspondence between the 
fraction of the $r_{500}$ covered by the aperture and the group surface 
brightness profile, as described in \citet{finoguenov07}. 
For most groups, the detection of their X-ray emission extends to $r_{500}$ and 
no correction for the aperture was made, or the applied corrections were 
minimal.
This allows us to link the observed properties of these groups, to groups detected at 
similar depths in the COSMOS field \citep{finoguenov07,scoville} and
whose X-ray luminosities were calculated in exactly the same fashion.
These COSMOS X-ray groups were binned by $L_{X}$ and redshift, and
average total halo masses ($M_{200}$) derived by \citet{leauthaud} for the sub-samples by stacked
weak gravitational lensing, producing a $M_{200}{-}L_{X}$ scaling
relation well described by a single power law. 

We estimate the total $M_{200}$ masses of our X-ray groups using the
$M_{200}{-}L_{X}$ relation of \citet{leauthaud} derived by performing a joint fit between the stacked COSMOS X-ray
groups and ten high-mass clusters from LoCuSS with analogous X-ray and
weak-lensing data (all ten are within our sample of 23). The resulting single power law relation
\begin{equation} \label{m200_lx}
\frac{\langle M_{200} E(z) \rangle}{M_{0}} = A \left( \frac{\langle
    L_{X} E(z)^{-1} \rangle}{L_{X,0}} \right)^{\alpha}
\end{equation}  
with $M_{0}{=}10^{13.70}\,h_{72}^{-1}\,{\rm M}_{\odot}$,
$L_{X,0}{=}10^{42.70}\,h_{72}^{-2}$\,erg\,s$^{-1}$, power-law index 
$\alpha{=}0.64{\pm}0.03$ and log$_{10}(A){=}0.03{\pm}0.06$, 
holds over two decades in mass,
$M_{200}{\sim}10^{13.5}{-}10^{15.5}\,{\rm M}_{\odot}$. \citet{allevato}
find that the observed bias of these COSMOS X-ray groups as measured
through the projected auto-correlation function is consistent 
with that predicted from the group masses derived via the above
relation. 
This relation produces group masses which are always within 20\% of those
resulting from the $M{-}L_{X}$ scaling relation of \citet*{lovisari},
whose {\em XMM-Newton} analysis derived total masses for 20 local groups assuming
hydrostatic equilibrium. The mass scatter at fixed $L_{X}$ is expected
to be 0.15\,dex \citep{kettula,lovisari}.

\begin{figure*}
\centerline{\includegraphics[width=180mm]{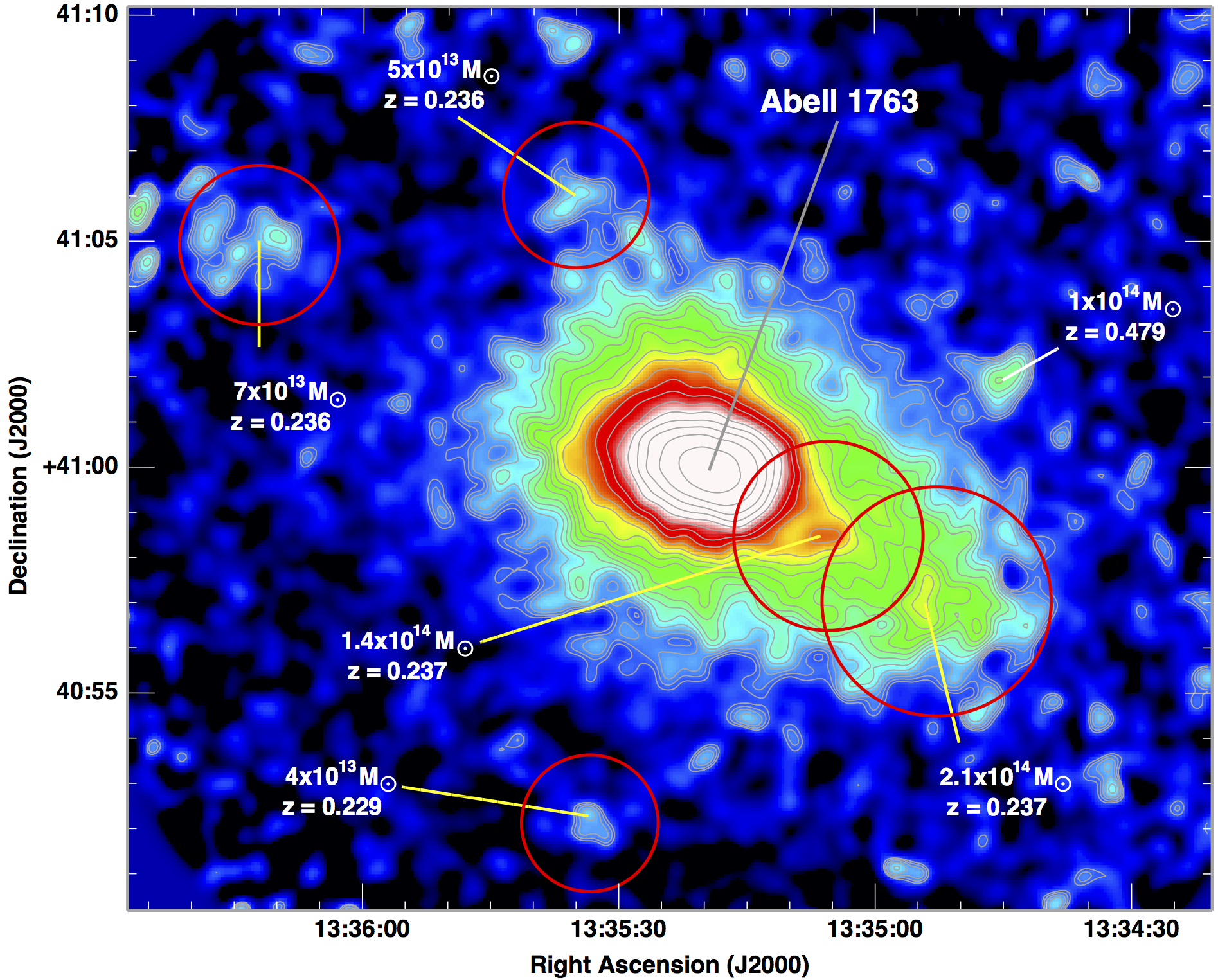}}
\caption{Extended X-ray emission from Abell 1763. All
  spectroscopically confirmed X-ray groups are indicated and labelled
  by their redshift and estimated M$_{200}$ value. Those groups which
  are infalling into Abell 1763 are marked by red circles of diameter $r_{200}$.} 
\label{a1763map}
\end{figure*}

\begin{figure*}
\centerline{\includegraphics[width=180mm]{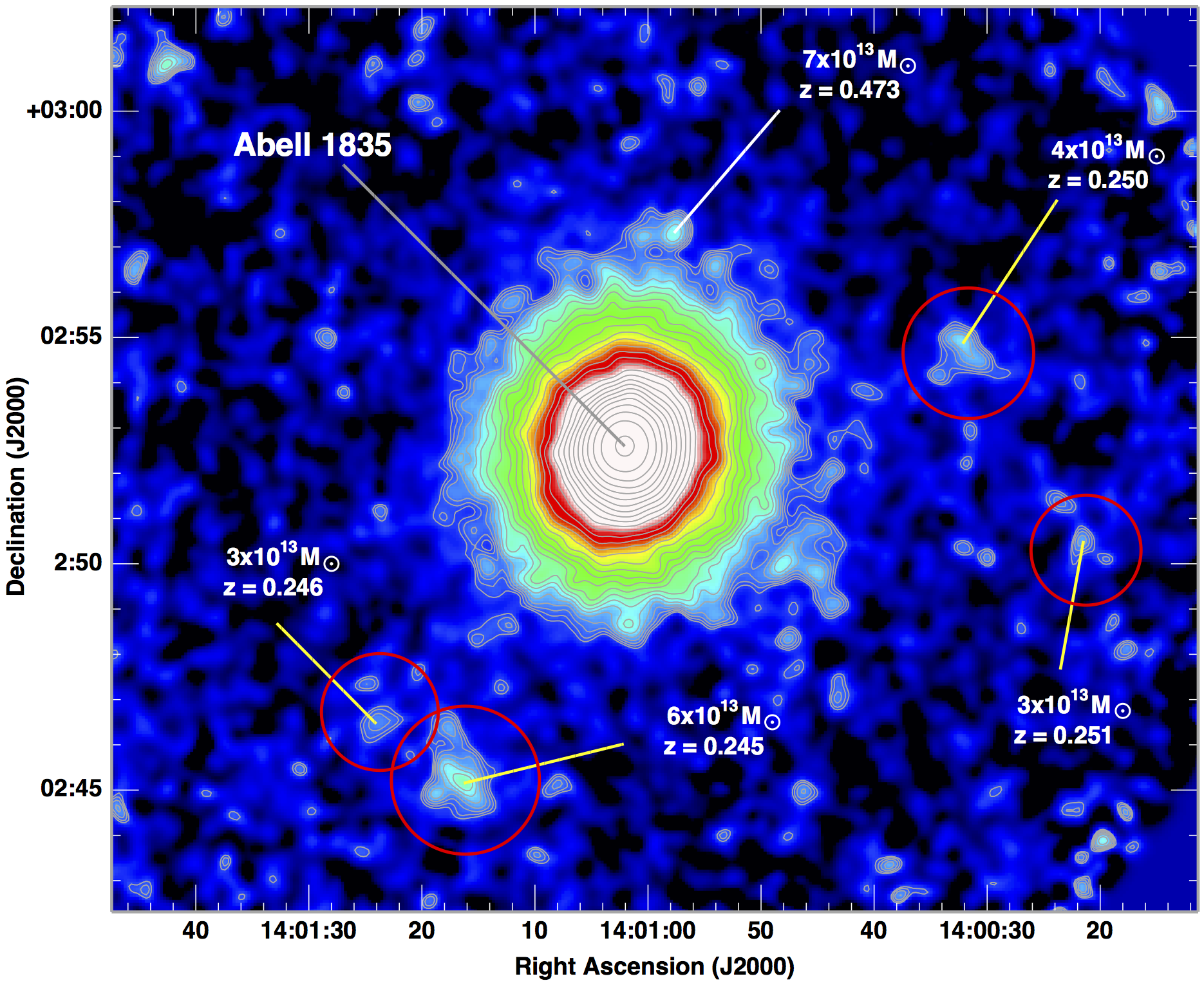}}
\caption{Extended X-ray emission from Abell 1835. All
  spectroscopically confirmed X-ray groups are indicated and labelled
  by their redshift and estimated M$_{200}$ value. Those groups which
  are infalling into Abell 1835 are marked by red circles of diameter $r_{200}$.} 
\label{a1835map}
\end{figure*}


\subsection{A catalogue of X-ray groups falling into massive clusters}

 X-ray groups which are infalling into the primary clusters are
identified from their projected cluster-centric radii and redshifts as
those located within the ``trumpet''-shaped caustic profile enclosing 
those galaxies identified as cluster members by \citet{haines13}.
In total 39 of our X-ray groups were identified as being within the
caustics of the primary cluster. 
The full list of infalling X-ray groups is presented in
Table~\ref{grouplist}, including their positions, redshifts, X-ray
luminosities, mass estimates, whether they have an obvious central 
dominant galaxy (Brightest Group Galaxy or BGG) or not, and the number of spectroscopically-confirmed members.
These groups all have at least one spectroscopic member, and indeed
31/39 have ${\ge}7$ members. The median number of spectroscopic
members is nine. The velocity dispersions of these groups, as
estimated using the gapper method \citep{beers}, lie in the range 150--650\,km/s$^{-1}$.

\begin{figure}
\centerline{\includegraphics[width=84mm]{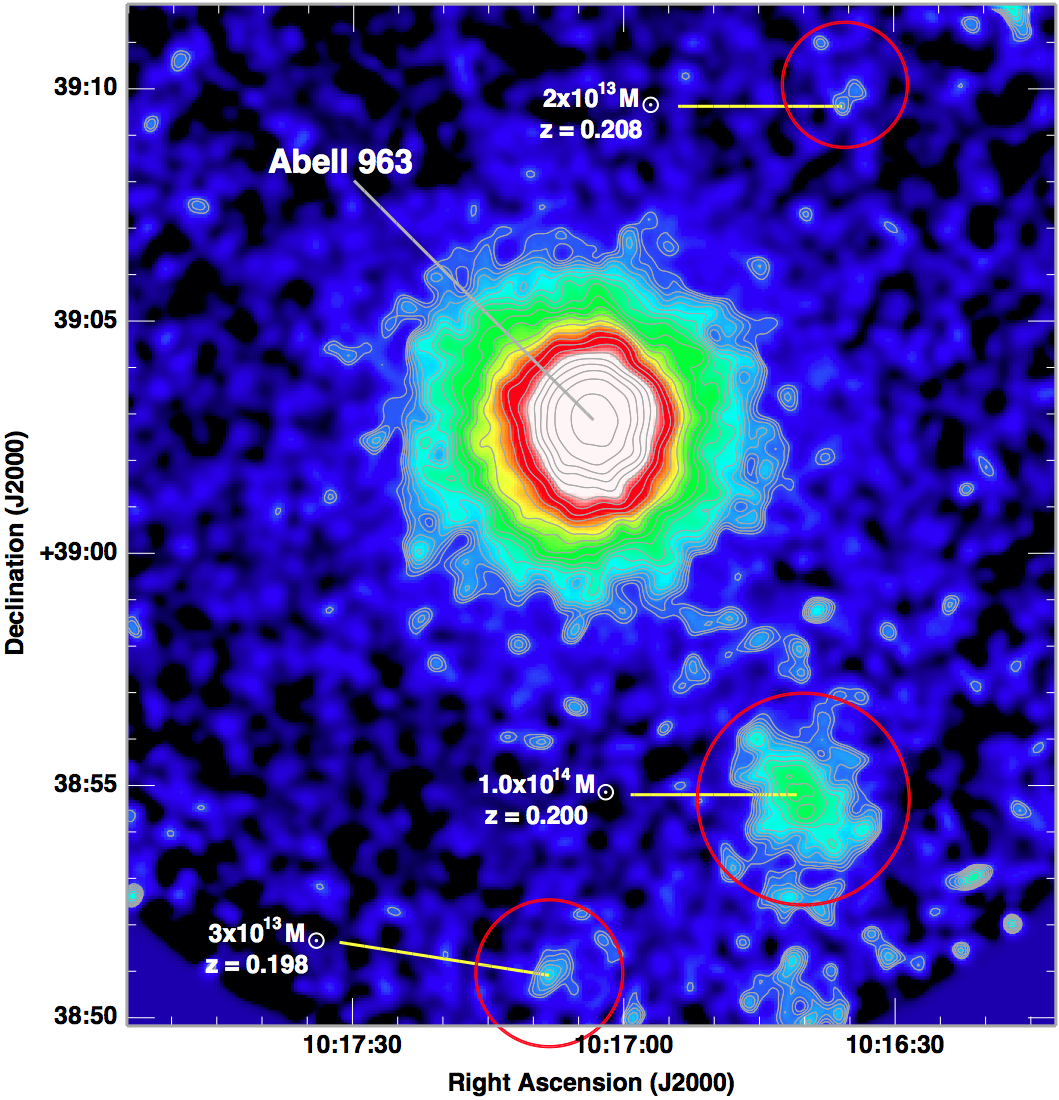}}
\caption{Extended X-ray emission from Abell 963. The three
  spectroscopically confirmed X-ray groups infalling into A963 are
  marked by red circles and labelled as in Figs.~\ref{a1763map}--\ref{a1835map}.} 
\label{a963map}
\end{figure}

\begin{figure}
\centerline{\includegraphics[width=84mm]{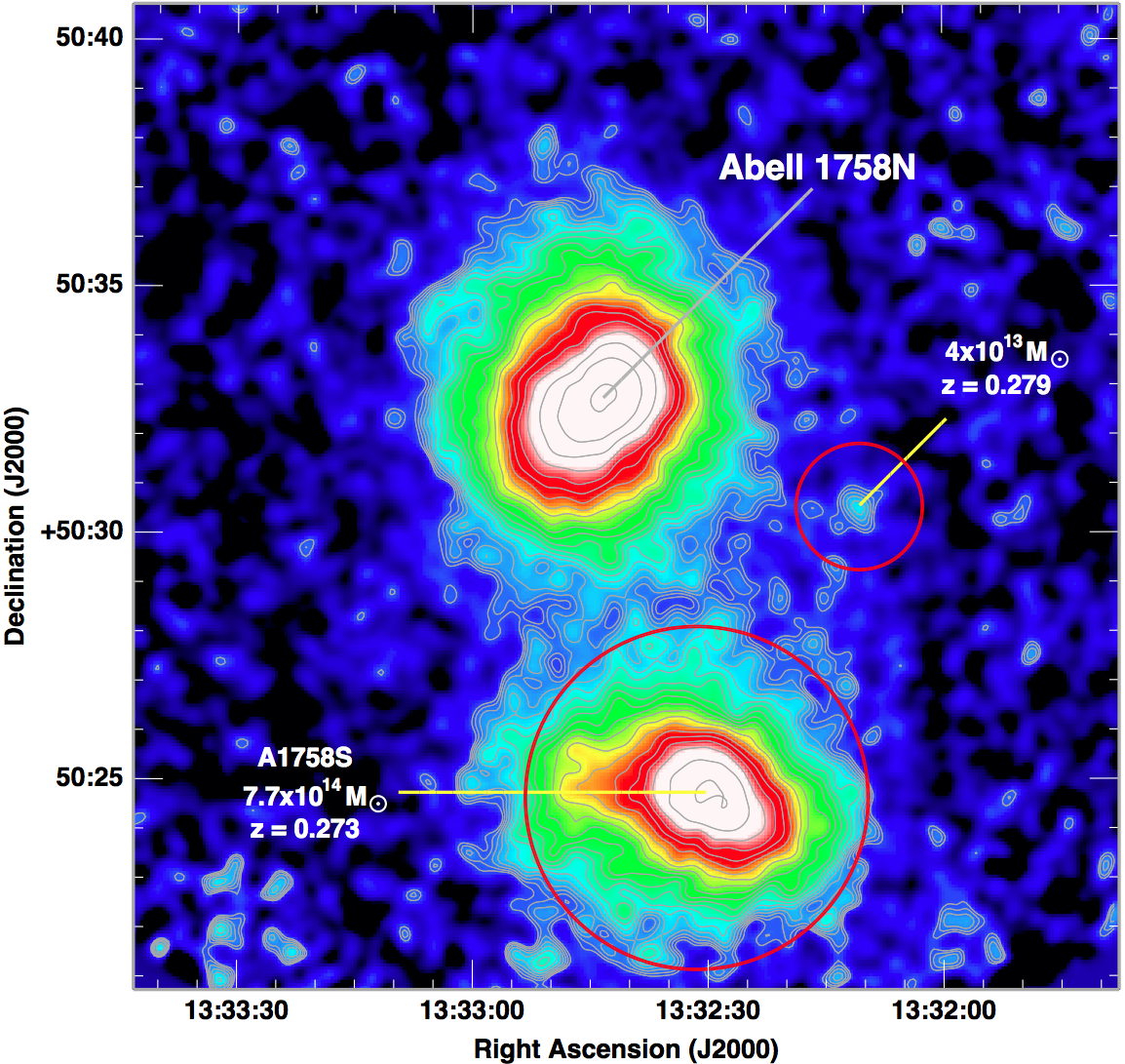}}
\caption{Extended X-ray emission from Abell 1758N. Both A1758S and the
  second X-ray group falling into A1758N are marked by red circles and
  labelled as before.}  
\label{a1758map}
\end{figure}

Figures~\ref{a1763map}--\ref{a1758map} show the 0.5--2.0\,keV band X-ray images of
Abell 1763, Abell 1835, Abell 963 and Abell 1758N, smoothed with a Gaussian kernel of width 12$^{\prime\prime}$ after removing point sources.
These enable the extended X-ray emission from groups in the outskirts
of the cluster and in the background to be identified. Each
X-ray detected group with confirmed redshift is labelled along with
its M$_{200}$ mass estimate. 

The X-ray emission from Abell 1763 extends significantly in the WSW
direction \citep[as seen previously by][Fig.~B.6]{zhang07}, including two sub-peaks that are identified as
${\sim}10^{14\,}{\rm M}_{\odot}$ groups. Both of these peaks are close
to extremely massive passive galaxies with
$\mathcal{M}{\sim}10^{11.6}{\rm M}_{\odot}$, featureless
bulges and the extended diffuse envelopes characteristic of BCGs. 
Two more groups with masses 5--7${\times}10^{13\,}{\rm M}_{\odot}$ lie
along the same axis, but in the opposite direction (ENE), notably
directly towards Abell 1770, some 13\,Mpc distant, and within the
filament of star-forming galaxies previously detected as feeding Abell
1763 along this axis \citep{fadda,edwards}. Again the centres of X-ray
emission from both groups are located close to massive passive galaxies.

Abell 1835 is a classic relaxed cool-core cluster, as demonstrated by
its regular circular surface brightness contours, and the most
luminous cluster in the ROSAT Brightest Cluster Sample \citep{ebeling98}. At larger radii
(0.5--1.0\,$r_{200}$) four X-ray groups are identified with masses
2.7--6.$1{\times}10^{13\,}{\rm M}_{\odot}$. The two most massive
groups were previously identified in the {\em XMM} data by
\citet{pereira} and in {\em Chandra} data by \citet{bonamente}, who
measured temperatures of 2.7 and 2.1\,keV for the pair.  
The X-ray peaks are centred on massive passive
galaxies with $\mathcal{M}{\sim}10^{11.3}{\rm M}_{\odot}$. The finding
of numerous groups in the infall regions of A1835 can explain its 
velocity dispersion profile which remains flat at 1\,500\,km\,s$^{-1}$
out to 2\,Mpc and significant substructures detected by the
Dressler-Shectman test \citep{czoske}. It also demonstrates how the
core of a cluster may behave as a relaxed system \citep{smith05}, while on larger
scales it is strongly disturbed due to the presence of infalling
groups. 

\begin{figure}
\centerline{\includegraphics[width=84mm]{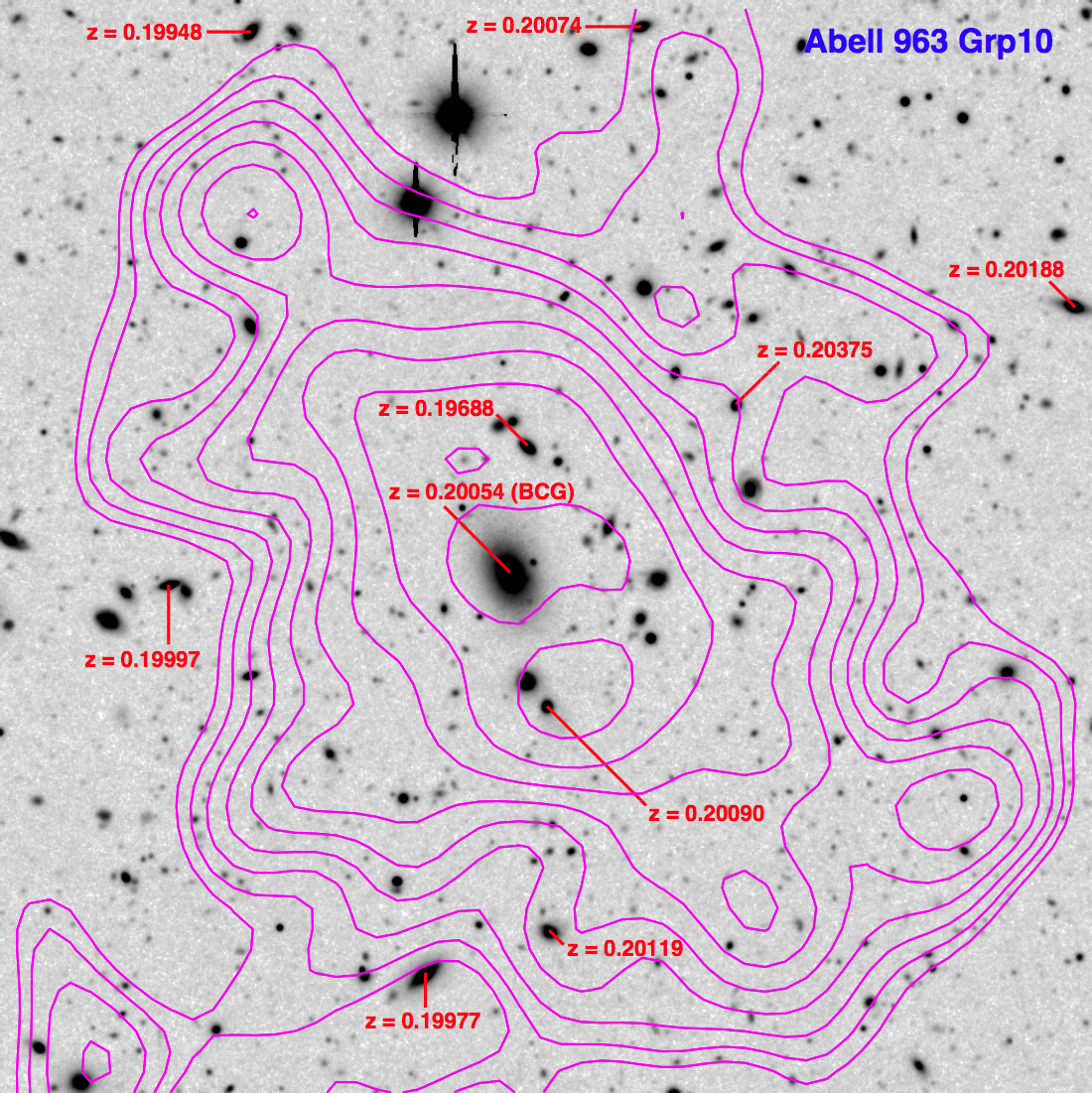}}
\caption{A $214^{\prime\prime}{\times}214^{\prime\prime}$ Subaru/SuprimeCAM $I_{C}$-band image centred on the infalling group
  A963-g10 ($M_{200}{=}1.0{\times}10^{14\,}{\rm M}_{\odot}$). Magenta contours indicate the extended X-ray emission
  from the group. Spectroscopic group members are indicated in red,
  and their redshift reported.}  
\label{a963grp10}
\end{figure}

\begin{figure}
\centerline{\includegraphics[width=84mm]{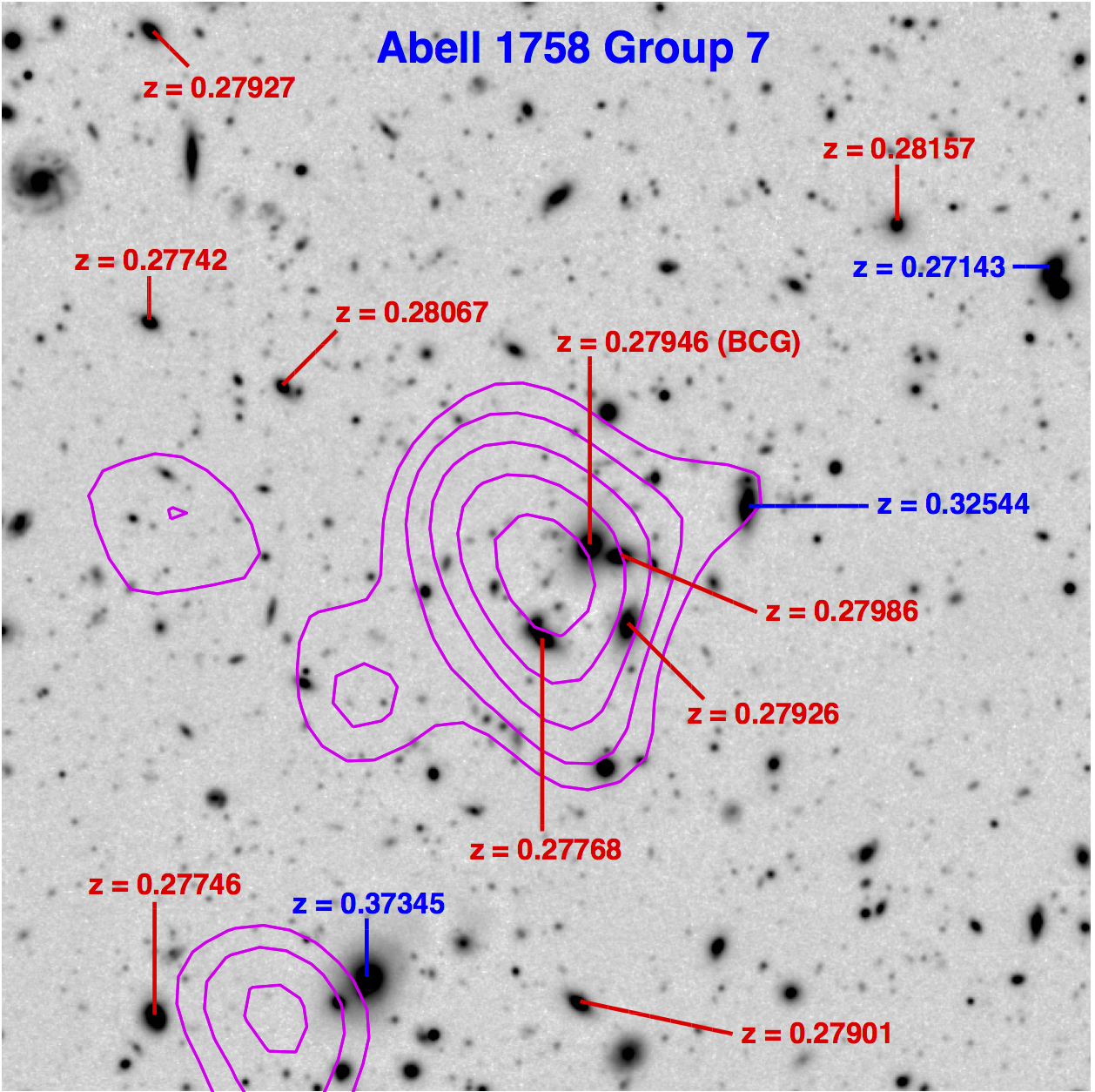}}
\caption{A $160^{\prime\prime}{\times}160^{\prime\prime}$ Subaru/SuprimeCAM $R_{C}$-band image centred on the infalling group
  A1758-g7 ($M_{200}{=}4.1{\times}10^{13\,}{\rm M}_{\odot}$). Magenta contours indicate the extended X-ray emission
  from the group. Spectroscopic group members are indicated in red,
  and other galaxies with known redshifts indicated in blue.}  
\label{a1758grp7}
\end{figure}

Abell 963 has been classified as a relaxed cluster based on the joint
HST strong-lensing and X-ray analysis of \citet{smith05}, although it
lacks the strong cool core of A1835. The {\em XMM} maps again reveal
significant sub-structure on large scales, with three infalling groups
identified. While two are relatively poor systems
($M_{200}{\sim}$2--3${\times}10^{13\,}{\rm M}_{\odot}$, the third is a
massive group with $M_{200}{=}1.0{\times}10^{14\,}{\rm M}_{\odot}$,
located at a projected distance of 1.0\,$r_{200}$ from A963. 
A Subaru $I_{C}$ image centred on the group is shown in Fig.~\ref{a963grp10}
 with the known group members labelled. The X-ray
 emission ({\em magenta contours}) is centred on a massive passive galaxy 
($\mathcal{M}{\sim}10^{11.2}{\rm M}_{\odot}$) which dominates the group.
The BUDHIES team
carried out an ultra-deep H{\sc i} survey of Abell 963 and its
environs, revealing that galaxies within the most massive X-ray group
are strongly deficient in H{\sc i}, relative to the lower mass groups
and the infall regions of the cluster, and providing evidence of the
impact of pre-processing on galaxies in infalling groups, stripping
their gas contents and quenching star-formation before they are 
accreted into the cluster \citep{jaffe}.

Abell 1758 is a well known double cluster system with two distinct
clusters (A1758N and A1758S) separated by 8\,arcmin \citep{rizza} in
the plane of the sky (or 2\,Mpc at $z{=}0.279$). Both A1758N and
A1758S are undergoing major mergers, with {\em Chandra} imaging of
A1758N revealing two remant cores separated by 800\,kpc and shock
fronts \citep{david}. Our {\em XMM} analysis has revealed a further sub-structure
within this complex system, a $4{\times}10^{13\,}{\rm M}_{\odot}$
group that lies 6\,arcmin (1.5\,Mpc) ESE from A1758N and at the same
redshift ($z{=}0.279$). The X-ray emission is centred on a compact clump of
four group members, including the BCG (Fig~\ref{a1758grp7}), with
many more group galaxies in the vicinity. 

\begin{figure}
\centerline{\includegraphics[width=84mm]{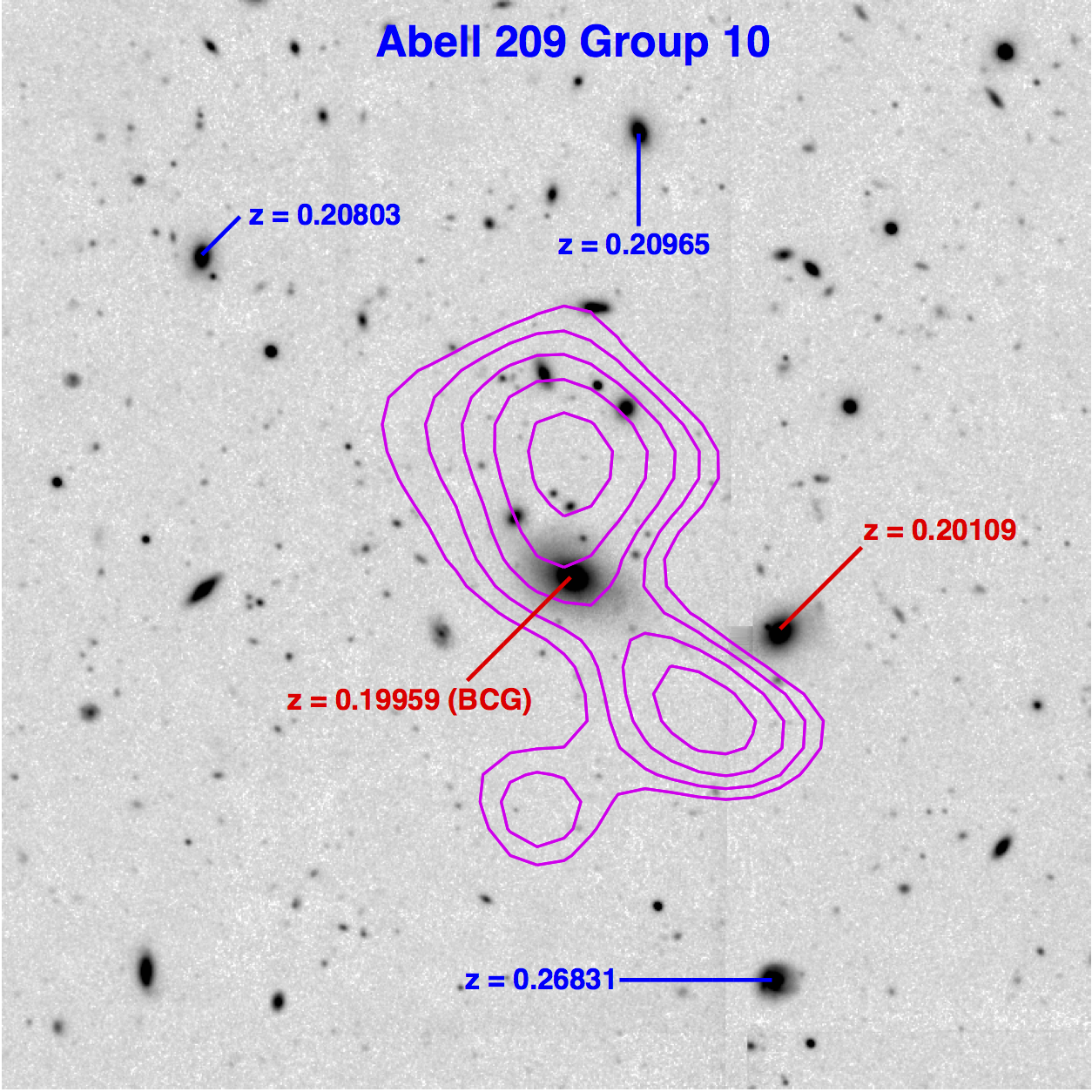}}
\caption{A $160^{\prime\prime}{\times}160^{\prime\prime}$ SuprimeCAM $i$-band image centred on the infalling group
  A209-g10 ($M_{200}{=}3.0{\times}10^{13\,}{\rm M}_{\odot}$). The two
  group members are indicated in red. Other galaxies with known
  redshifts are labelled in blue.}  
\label{a209grp10}
\end{figure}

\begin{figure}
\centerline{\includegraphics[width=84mm]{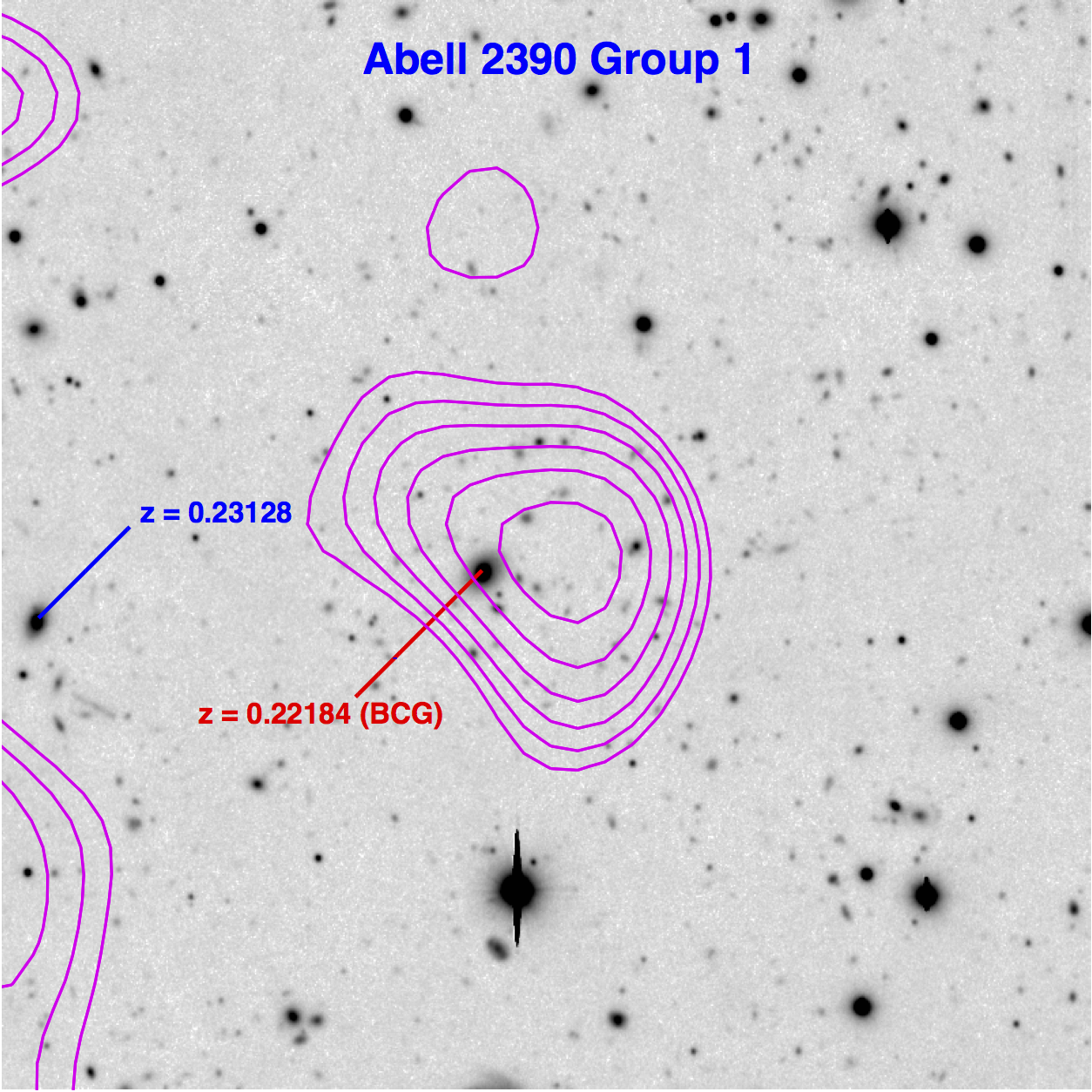}}
\caption{A $160^{\prime\prime}{\times}160^{\prime\prime}$ SuprimeCAM $R_{C}$-band image centred on the infalling group
  A2390-g1 ($M_{200}{=}4.5{\times}10^{13\,}{\rm M}_{\odot}$).}  
\label{a2390grp1}
\end{figure}

Figures~\ref{a209grp10} and~\ref{a2390grp1} show two examples of our
poorest infalling X-ray groups, A209-g10 and A2390-g1. In the former,
the X-ray emission is centred on an obvious dominant group galaxy ($\mathcal{M}{\sim}10^{11.2\,}{\rm M}_{\odot}$),
with a nearby second bright galaxy within 400\,km\,s$^{-1}$. In the
latter, there is only one bright galaxy located within the X-ray
contours, a massive ($\mathcal{M}{\sim}10^{11.0\,}{\rm M}_{\odot}$)
passive galaxy. In both cases, the association of the X-ray emission
with the group galaxies appears robust and unambigious.

\subsection{A comparison sample of galaxy groups around clusters in
  the Millennium simulation}
\label{millennium}

To understand the fates of the infalling X-ray groups, estimate the mass
completeness of the {\em XMM} observations and to compare our results to
predictions from $\Lambda$CDM cosmological models, we have created a
comparison sample of galaxy groups in the vicinity of the 75 most
massive clusters (M$_{200}{>}4.0{\times}10^{14}h^{-1}{\rm M}_{\odot}$
at $z{=}0.0$) from the Millennium simulation
\citep{springel}, a cosmological dark matter simulation covering a
($500\,h^{-1}$\,Mpc)$^{3}$ volume. 
Following \citet{haines15} we have extracted dark
matter halos with $M_{200}{>}10^{13}h^{-1}\,{\rm M}_{\odot}$ from the
MPA Halo (MHalo) catalogue within $20{\times}20{\times}140\,h^{-3}\,{\rm
  Mpc}^{3}$ volumes centred on each cluster. These volumes are
extended in the $z$-direction so that, for a distant observer viewing
along this axis, all galaxy groups with line-of-sight (LOS) velocities within
5000\,km\,s$^{-1}$ of the cluster redshift are included, enabling
projection effects to be fully accounted for and quantified. The group
halo positions and velocities relative to the primary cluster halo are
measured at the $z{=}0.21$ snapshot from the simulation, and
artificial observations created assuming a distant observer along the $z$-axis. 
Those DM halos whose LOS velocities place them within the caustics
defined by the cluster galaxy members used in \citet{haines15} are
then retained to form our comparison sample of simulated infalling
galaxy groups, and the $M_{200}{-}L_{X}$ scaling-relation of Eq.~\ref{m200_lx} used to predict their X-ray luminosities. 

The completeness of each {\em XMM} image as a function of group mass is
measured following the procedure outlined in \citet{finoguenov15}, by
generating simulated galaxy groups of a given mass and redshift, and the tabulation of \citet{finoguenov07} to predict the parameters
of the beta model used to describe the surface brightness profile of
their X-ray emission. Unlike \citet{finoguenov15} we use the simulated
infalling groups from the Millennium simulation to model the mass and
radial distribution of groups around the cluster, rather than assume a
random spatial distribution and mass distribution defined by a
$\Lambda$CDM cosmological model. This step is necessary because the
distribution of groups in and around clusters is not described by  
linear growth theory and simulations are required. The particular choice 
of the cosmology in the simulations is not so important, as the 
abundance of subhalos is a not a very sensitive function of the 
cosmology \citep{taylor05} and the subhalo mass function scales well at any redshift as a 
function of the subhalo ratio to the total halo mass \citep{giocoli08}.

\begin{figure*}
\centerline{\includegraphics[width=180mm]{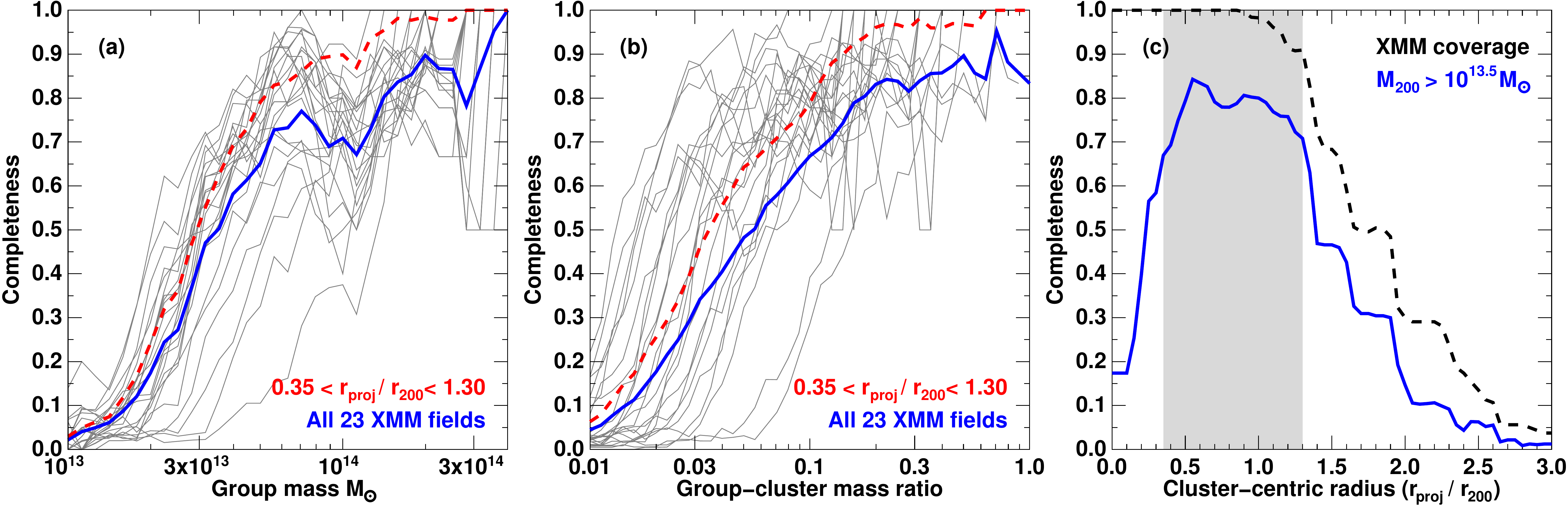}}
\caption{Fraction of simulated galaxy groups recovered by the wavelet-reconstruction
detection algorithm in each of the XMM images (gray lines) as a
function of group $M_{200}$ mass ({\em left panel}) and group-cluster
mass ratio ({\em central panel}). The blue curves shows the mean recovery rate
averaged over all 23 clusters. The red dashed curves shows the mean
recovery rate after excluding the cluster core regions with
$r_{proj}{<}0.35\,r_{200}$ and the outer regions affected by
vignetting ($r_{proj}{>}1.3\,r_{200}$). ({\em right panel}) Fraction of
simulated galaxy groups with $M_{200}{>}10^{13.5}\,M_{\odot}$ covered
by the XMM imaging (black dashed curve) and recovered by the detection
algorithm (blue curve) as a function of cluster-centric radius. The
gray shaded region indicates the radial range over which the
recoverate rate of X-ray groups within the {\em XMM} images should be
highest (red dashed lines in panels a,b).}
\label{completeness}
\end{figure*}

 The simulated infalling X-ray groups are placed into each of the {\em XMM}
images one at a time, and the wavelet detection algorithm applied,
producing catalogues of detected simulated groups for each {\em XMM}
cluster observation.    
Figure~\ref{completeness} plots the fraction of these simulated groups recovered by the
wavelet-detection algorithm as a function of group halo mass 
(M$_{\rm 200}$; Fig.~\ref{completeness}a), group-cluster mass ratio
(Fig.~\ref{completeness}b) and cluster-centric radius
(Fig.~\ref{completeness}c). The blue and grey curves respectively show the recovery
rates averaged over the 23 clusters, and for each individual {\em XMM}
observation.
We expect to detect ${\sim}7$0\% of groups with M$_{\rm
  200}{\sim}5{\times}10^{13\,}{\rm M}_{\odot}$ over the 23 {\em XMM}
fields. The recovery rate only rises slightly to higher masses, which
is partly due to some of the {\em XMM} observations having high
backgrounds, negatively affecting the detection rate even at masses
approaching $10^{14\,}{\rm M}_{\odot}$. 

 The key cause of incompleteness at masses above $10^{13.5\,}{\rm M}_{\odot}$
is revealed in Fig.~\ref{completeness}c, where the recovery rate ({\em
  blue curve})
plummets from ${\sim}80$\% at $0.5{<}(r_{proj}/r_{200}){<}1.0$ to just ${\sim}2$0\%
in the cluster cores ($r_{proj}{\la}\,0.35\,r_{200}$). This is due to
difficulty in distinguishing the X-ray emission from groups from the
much larger emission from the primary cluster. 
This explains the notable absence of X-ray groups in the cluster core
regions in our sample (Table~\ref{grouplist}; Fig.~\ref{zspace}). 
This affects our overall mass completeness level, and the red dashed
lines in Figs~\ref{completeness}a,b shows the improved completeness
levels after excising the cluster cores ($r_{proj}{<}0.35\,r_{200}$).

We also start missing X-ray groups at large cluster-centric radii due to the limited
field-of-view of the {\em XMM} instruments, as shown by the black dashed
curve in Fig.~\ref{completeness}c. While the {\em XMM} data provides
complete coverage inside $r_{\rm 200}$, the coverage fraction drops
rapidly beyond 1.3\,$r_{200}$.
Our {\em XMM} data is thus most efficient at detecting infalling X-ray groups at cluster-centric distances
of 0.35--1.3\,$r_{200}$.


\section{Results}

A total of 39 X-ray groups are identified across the 23 {\em XMM}
images as being associated with the primary clusters, down to a
signal-to-noise limit of 3. Six of the
clusters (A267, A291, A383, A1689, RXJ2129, Z2089) have no X-ray detected groups in their infall regions, while
Abell 1763 has the most with five. The numbers of groups around each
cluster are consistent with the 39 groups being allocated
randomly to the 23 clusters.

 
\subsection{Spatial and velocity distribution of the infalling X-ray groups}

\begin{figure*}
\centerline{\includegraphics[width=120mm]{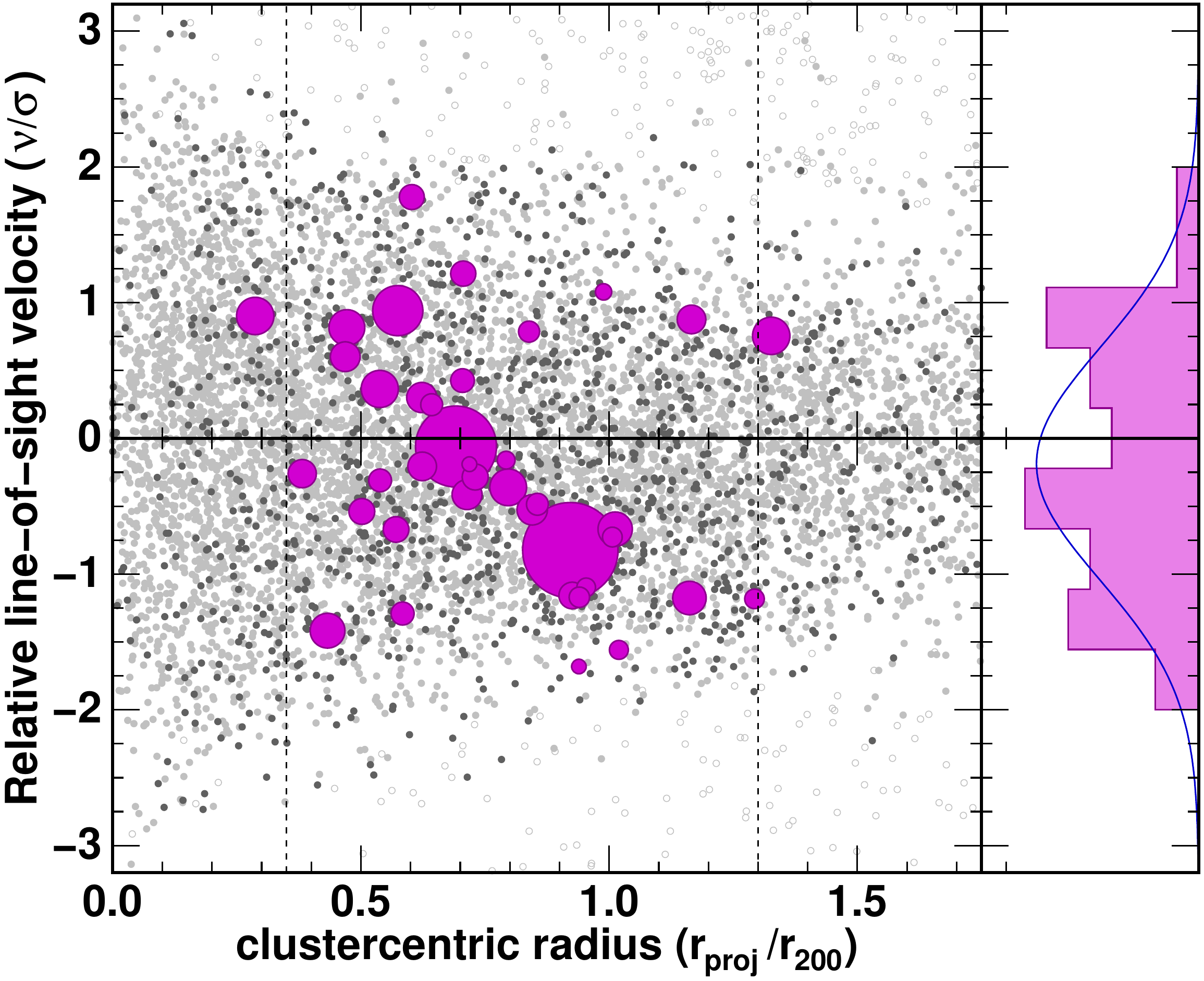}}
\caption{Stacked phase-space
  diagram, \mbox{$(\nu_{los}{-}\langle\nu\rangle)/\sigma_{\nu}$} versus $r_{proj}/r_{200}$
 of the 39 XMM groups ({\em magenta symbols}) and member galaxies
 ({\em grey solid points}) for all 23 clusters in our sample. The size of each symbol indicates the
 group mass. Darker grey symbols indicating star-forming galaxies
 detected at 24$\mu$m. Open symbols indicate field galaxies. The
 vertical dashed lines enclose the $0.35{\le}(r_{proj}/r_{200}){<}1.30$ region
 where the {\em XMM} data is most complete.
The
 solid histogram on the right shows the distribution of relative LOS
 velocities for the 39 XMM groups, while the blue curve shows
 a Gaussian distribution with the same mean and standard deviation.} 
\label{zspace}
\end{figure*}

Figure~\ref{zspace} shows the distribution of the 39 X-ray groups ({\em
  magenta symbols}) in the
stacked caustic diagram. This plots the LOS velocity of each 
group relative to the central redshift of the primary cluster, 
scaled by the velocity dispersion of all cluster members within
$r_{200}$ \citep[$\sigma_{\nu_{cl}}$; taken
from][]{haines15}, against its projected cluster-centric distance. 
This shows how all these groups lie within the ``trumpet''-shaped region formed by
the galaxies ({\em solid grey points}) that have been spectroscopcially identified as members of
the same 23 clusters, demonstrating that the groups are indeed
associated with the clusters. 

The overall distribution of relative LOS velocities for these 39 groups is
shown by the histogram on the right. As discussed in detail in
\citet{haines15}, both the width and shape of the LOS velocity distribution
of populations of objects in and around galaxy clusters depend
strongly on when they have been (or will be) accreted into the
cluster. Low LOS velocity dispersions and Gaussian distributions are
indicators of virialized populations, while high LOS velocity
dispersions and flat top-hat distributions are associated with objects
on their first infall \citep{haines15,hikage}. 

While the velocity dispersion of the X-ray groups about the 
cluster redshift is marginally lower than that of the overall cluster
galaxy population,
$\sigma(\nu_{gr}{-}\nu_{cl}){=}0.86{\pm}0.08\,\sigma_{\nu_{cl}}$, 
half of the XMM groups are located along the caustics where objects
on their first infall into the clusters are expected to be found. The
histogram shows an excess of groups with velocities around 
${-}1.4\,\sigma_{\nu_{cl}}$ and ${+}1.0\,\sigma_{\nu_{cl}}$,
relative to expectations from a Gaussian distribution ({\em blue dashed
curve}), and a shortfall of groups with LOS velocities around zero. 
The
kurtosis of the group-cluster LOS velocity distribution is negative
($\gamma{=}{-}0.82{\pm}0.36$), being inconsistent at the $2.3\sigma$ level 
with that of a Gaussian distribution ($\gamma{=}0.0$), and closer to the value expected for a flat top-hat
distribution ($\gamma{=}{-}1.2$). 

The infalling X-ray groups are heavily concentrated within the
radial range 0.35--1.3\,$r_{200}$ (vertical dashed lines) where
the {\em XMM} data are predicted to be most complete
(Fig.~\ref{completeness}c), with 37/39 groups from our sample found
within this range.
That is not to say that we could not detect infalling X-ray groups beyond
1.3\,$r_{200}$. In fact, of the 52 other X-ray groups detected by {\em
  XMM}, but with redshifts inconsistent with that of the primary cluster, 26
were found at $r_{proj}{>}1.3\,r_{200}$. 

\begin{figure}
\centerline{\includegraphics[width=84mm]{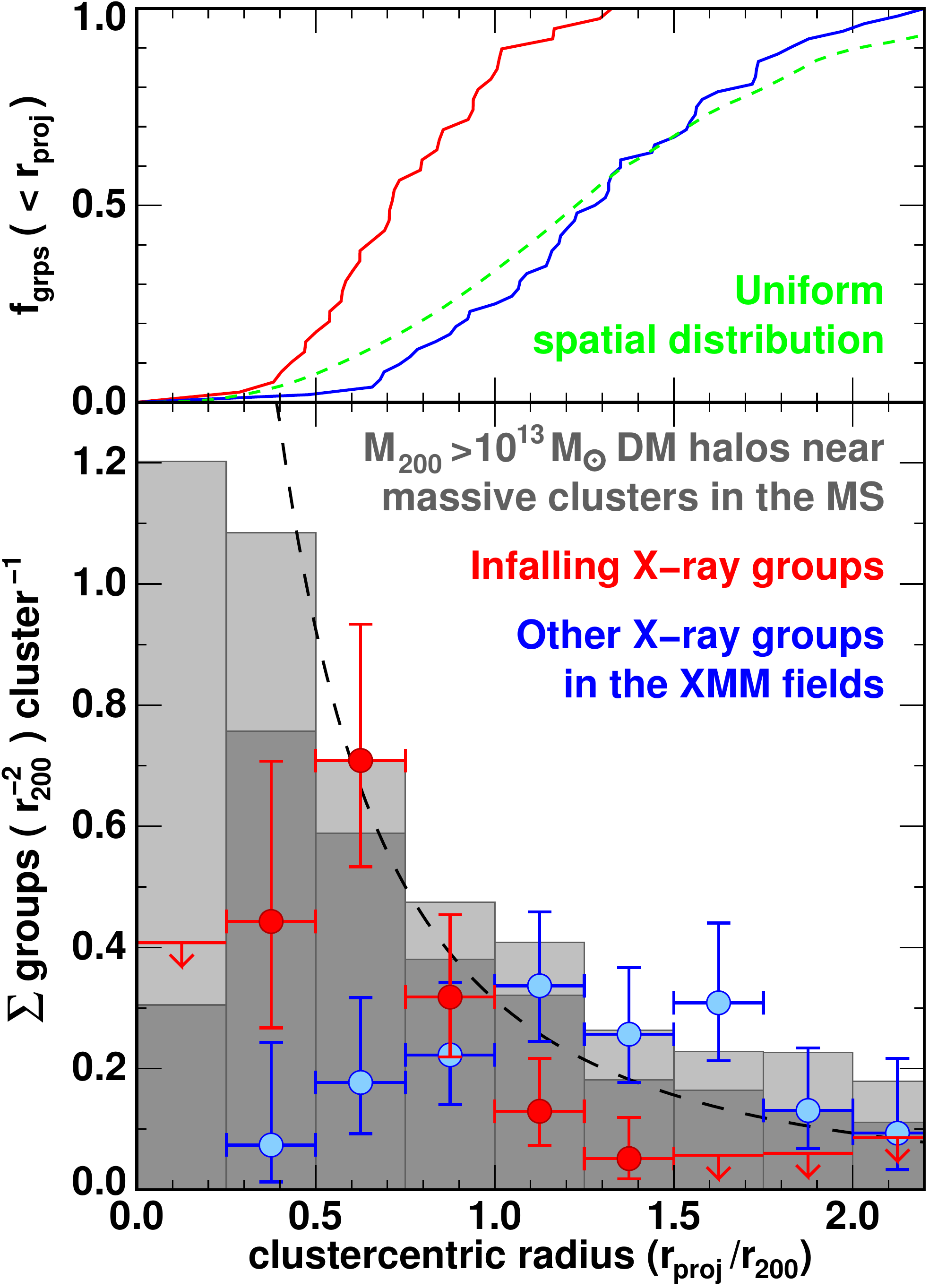}}
\caption{Composite surface number density distribution $\Sigma(r)$ of the
39 observed infalling X-ray groups ({\em red points}) and 52 other
``isolated'' field X-ray groups ({\em blue points}) as a function of
projected cluster-centric radius (lower panel). For those radial bins
containing no infalling groups, the Poisson 1$\sigma$ upper limit
\citep{gehrels} is shown.
The light grey shaded histrogram shows the predicted number density distribution
of M$_{200}{>}10^{13\,}{\rm M}_{\odot}$ groups as a function of
projected cluster-centric radius around the 75 most massive clusters
in the Millennium simulation at $z{=}0.21$. The dark shaded
distribution indicates the predicted radial distribution of those
groups which would be detected, applying the completeness correction
of Fig.~\ref{completeness}c. The dashed curve shows the best-fit
NFW profile to the radial distribution of cluster galaxies \citep[$c_{g}{=}3.01$;][]{haines15}.
The upper panel shows the cumulative
radial distribution $f({<}r_{proj})$ of the infalling X-ray groups
({\em red curve}) and ``isolated'' X-ray groups
detected in the same {\em XMM} fields. The dashed green curve indicates
  the predicted cumulative radial distribution if groups were
  uniformly distributed across the {\em XMM} images, taking into
  account the effects of incompleteness in the cluster cores
  (Fig.~\ref{completeness}c).}
\label{radial}
\end{figure}

Figure~\ref{radial} compares the surface number density $\Sigma(r)$
distributions of the 39 infalling X-ray groups ({\em red points}) and the other 52
``isolated'' X-ray groups (i.e. not in the vicinity of a massive cluster; {\em blue points}) found in the same {\em
  XMM} fields, as a function of projected cluster-centric radius. 
The two radial distributions are markedly
different. The infalling X-ray groups show a sharp peak at
0.5--0.75\,$r_{200}$, before dropping off rapidly at larger radii and
no infalling groups beyond 1.33\,$r_{200}$, 
while the other X-ray groups in the same field show a rather flat
radial distribution over 0.5--1.75\,$r_{200}$. In both cases, the
number densities of groups drops inside 0.5\,$r_{200}$ and are absent
within 0.25\,$r_{200}$, due to the inability to detect X-ray groups 
projected close to the cluster core in the {\em XMM} data (Fig.~\ref{completeness}c).
The upper panel shows the corresponding cumulative radial
distributions of the infalling X-ray groups ({\em red curve}) and
isolated ``field'' groups ({\em blue curve}). The infalling groups are found much closer on average
to the primary cluster than the back/foreground X-ray groups. The
median projected cluster-centric distance of the infalling groups is
0.71\,$r_{200}$, compared to 1.29\,$r_{200}$ for the remaining X-ray
groups detected in the same {\em XMM} fields. The largest difference
is seen at the virial radius, with 35/39 (90\%) infalling groups having
$r_{proj}{<}1.02\,r_{200}$, while 39/52 (75\%) of the isolated groups lie
at $r_{proj}{>}1.02\,r_{200}$. The non-parametric Mann-Whitney U-test
confirms the radial distributions of the infalling and field groups
to be inconsistent at the 6.0$\sigma$ level. 

As these other X-ray
groups are not associated with the cluster, we should expect them to
be uniformly distributed across the {\em XMM} images, but then be affected
by the same radial selection biases as the infalling groups (Fig.~\ref{completeness}c). 
The green dashed curve shows the expected cumulative radial
distribution of groups assuming a uniform spatial distribution over
the {\em XMM} images, and taking into account the loss of sensitivity
in the cluster core regions. The fore/background X-ray groups are
consistent with being randomly distributed across the {\em XMM}
fields, while the infalling X-ray groups are clearly not.

The strong preference of infalling X-ray groups to lie within 
$r_{200}$ can be understood in terms of the expected clustering
of group-mass systems around massive clusters. 
The light-grey shaded histogram shows the surface number density 
of M$_{200}{>}10^{13\,}{\rm M}_{\odot}$ DM halos (groups)
in the vicinity of the 75 most massive clusters from the Millennium
Simulation (MS), as a function of projected cluster-centric distance. 
The surface number density shows a sharp peak inside 0.5\,$r_{200}$,
before rapidly dropping to larger radii, falling sixfold by $r_{proj}{\sim}2\,r_{200}$.
This is the predicted radial distribution of groups before accounting
for observational biases, and the difficulty in detecting X-ray groups within 0.5\,$r_{200}$ above
the much greater emission from the cluster ICM pushes the expected 
peak out to 0.25--0.75\,$r_{200}$ ({\em darker histogram}). The sharp increase in the projected number
density of groups moving towards the cluster centre parallels that seen
also for the member galaxies \citep[Fig.~5]{haines15}, which was best fit by an
NFW profile with $c_{g}{=}3.01{\pm}0.16$ \citep[{\em dashed curve};][]{haines15}.
Such an NFW profile also describes well the predicted radial distribution
of groups ({\em grey histograms}) over 
0.5--2.2$\,r_{200}$, but is inconsisent with our lack of infalling X-ray
groups beyond $1.5\,r_{200}$. 
This largely reflects
the fact that clusters lie at the centres of large-scale
(${\ga}10$\,Mpc) overdensities
that extend well beyond the virial radius \citep{frenk}. These
large-scale overdensities are collapsing inwards towards the cluster,
dragging the infalling X-ray groups and galaxies with them \citep[Figs.~9,10]{haines15}. 


\subsection{The mass function of infalling X-ray groups}

\begin{figure}
\centerline{\includegraphics[height=72mm]{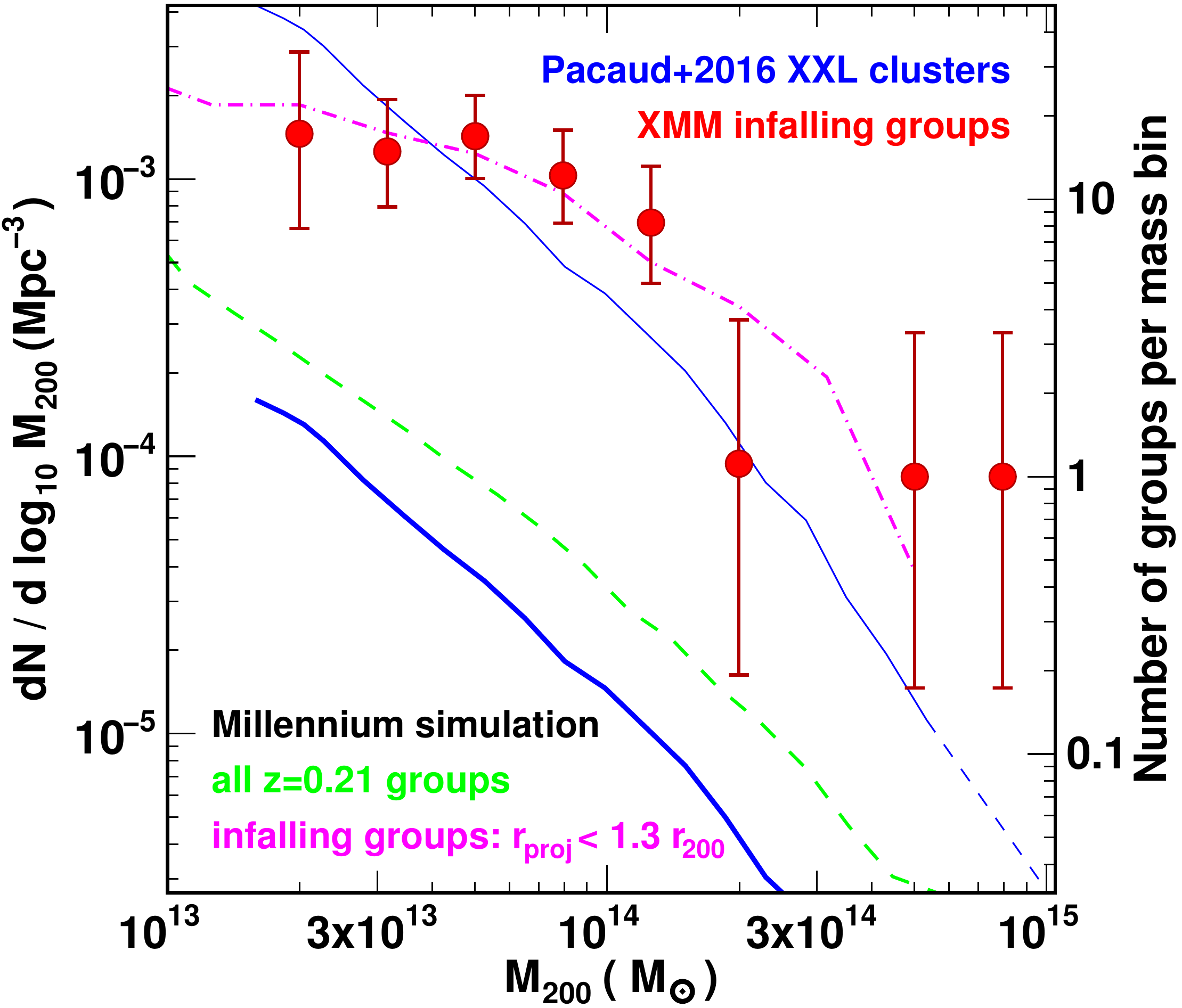}}
\caption{Mass function of the 39 {\em XMM}-detected groups with
  redshifts associating them with the 23 primary clusters (red solid 
  symbols). Error bars indicate Poisson uncertainties based on
  \citet{gehrels}. The thick blue curve indicates the mass
  function derived from the X-ray luminosity function of groups and clusters from the XXL survey
 \citep{pacaud}. The thin blue curve shows the same mass function,
 normalized upwards by a factor 26 (see text).
The dot-dashed magenta curve shows the mass function
  of simulated infalling galaxy groups within $1.3\,r_{200}$ (projected) of the
  75 most massive clusters in the Millennium simulation at $z{=}0.21$,
while the dashed green curve indicates the $z{=}0.21$ halo mass function averaged over
the whole Millennium simulation.}
\label{massfunction}
\end{figure} 

Figure~\ref{massfunction} shows the mass function (MF) of the
infalling X-ray groups
 ({\em red points}), after correcting for incompleteness
 (${\S}$\ref{millennium}). The left-hand axis shows the MF in units of
 groups per dex in mass per comoving Mpc$^{3}$. The
 comoving volume containing the infalling group sample for each
 cluster is estimated as that within the redshift limits
corresponding to the cluster caustics and extending over a circular area of sky of
radius $1.3\,r_{200}$. Summing these 23 volumes gives a grand total
comoving volume of $5.9{\times}10^{4\,}{\rm Mpc}^{3}$. 

The slope of the
mass function appears to flatten off for masses below
${\sim}10^{14\,}{\rm M}_{\odot}$. The shape and overall normalization
of the MF (in terms of groups per mass bin per cluster; right-hand
axis) is consistent with that predicted by the 
MF of infalling galaxy groups with
projected separations ${<}1.3\,r_{200}$ from the 75 most massive
clusters in the Millennium simulation ({\em dot-dashed magenta curve}). 

For comparison, the thick blue curve shows an estimate of
the overall MF of clusters in the universe at $z{\sim}0.2$, obtained by applying the
same $M_{200}-L_{X}$ relation to the X-ray luminosity function (XLF) of \citet{pacaud}. 
This XLF is based on a flux-limited sample of the 100 brightest extended X-ray sources
found in the XXL survey \citep{pierre}. This is the largest
programme carried out by {\em XMM-Newton}, covering a total area of
50\,deg$^2$ (412 XMM pointings) over two fields, to comparable depths
(10\,ksec) as those used here, with the objective of providing a statistical and
representative sample of groups and clusters out to $z{\sim}0.5$ (and above), suitable for
constraining cosmological parameters.
Most of these systems are located between $z{=}0.1$ and 0.5.
The X-ray luminosities of \citet{pacaud} were measured over the
0.5--2.0\,keV spectral band, and so were first divided by a global factor of
0.59 to k-correct them to the 0.1--2.4\,keV band used here
and in \citet{leauthaud}. 
The most notable difference between the
``cosmic'' MF and that of our
infalling X-ray groups, is the overall normalization. 
The comoving number density of X-ray groups in the infall regions of
clusters is more than an order of magnitude higher than that seen in
the XXL survey volume. The ``cosmic'' MF has to be normalized upwards by a factor 
${\sim}26$ ({\em thin blue curve}) in order to predict the same
overall number of $M_{200}{>}10^{13.2\,}{\rm M}_{\odot}$ groups as that observed in the XMM infalling
group sample.

Moreover, over the remainder of the $0.15{\le}z{<}0.30$ volume covered
by our 23 {\em XMM} images, we detect only ten more X-ray groups above
the $3\sigma$ SNR threshold, despite this volume being 5.76${\times}$
larger than that confined within the redshift limits of the
clusters. This corresponds to an over-abundance of X-ray groups in the
cluster infall regions of a factor ${\sim}22$, comparable to the
previous estimate, and confirming that the infall regions of
clusters are ${\sim}25{\times}$ overdense in group-mass systems with
respect to the cosmic average at that redshift. 
 
The shapes of the two MFs also appear different. 
The flattening seen in the MF of the infalling X-ray
groups is in marked contrast to the much steeper MF of XXL systems, which
can be well described as a single power law ($N(M){\propto}M^{-1.6}$) 
without any sign of a break. 
This steep, power-law form closely resembles the global MF of DM halos averaged over the
full volume of the Millennium simulation at $z{=}0.21$
({\em dashed green curve}), where no break in the MF is apparent. 
The clear difference in the mass functions of the X-ray groups
around massive clusters 
presented here and of X-ray groups sampled over a
large, representative volume of the Universe through the XXL survey,
reproduces well the predicted effect of the overdense cluster environment on the
mass function of DM halos seen within the Millennium simulation.

\begin{figure}
\centerline{\includegraphics[height=72mm]{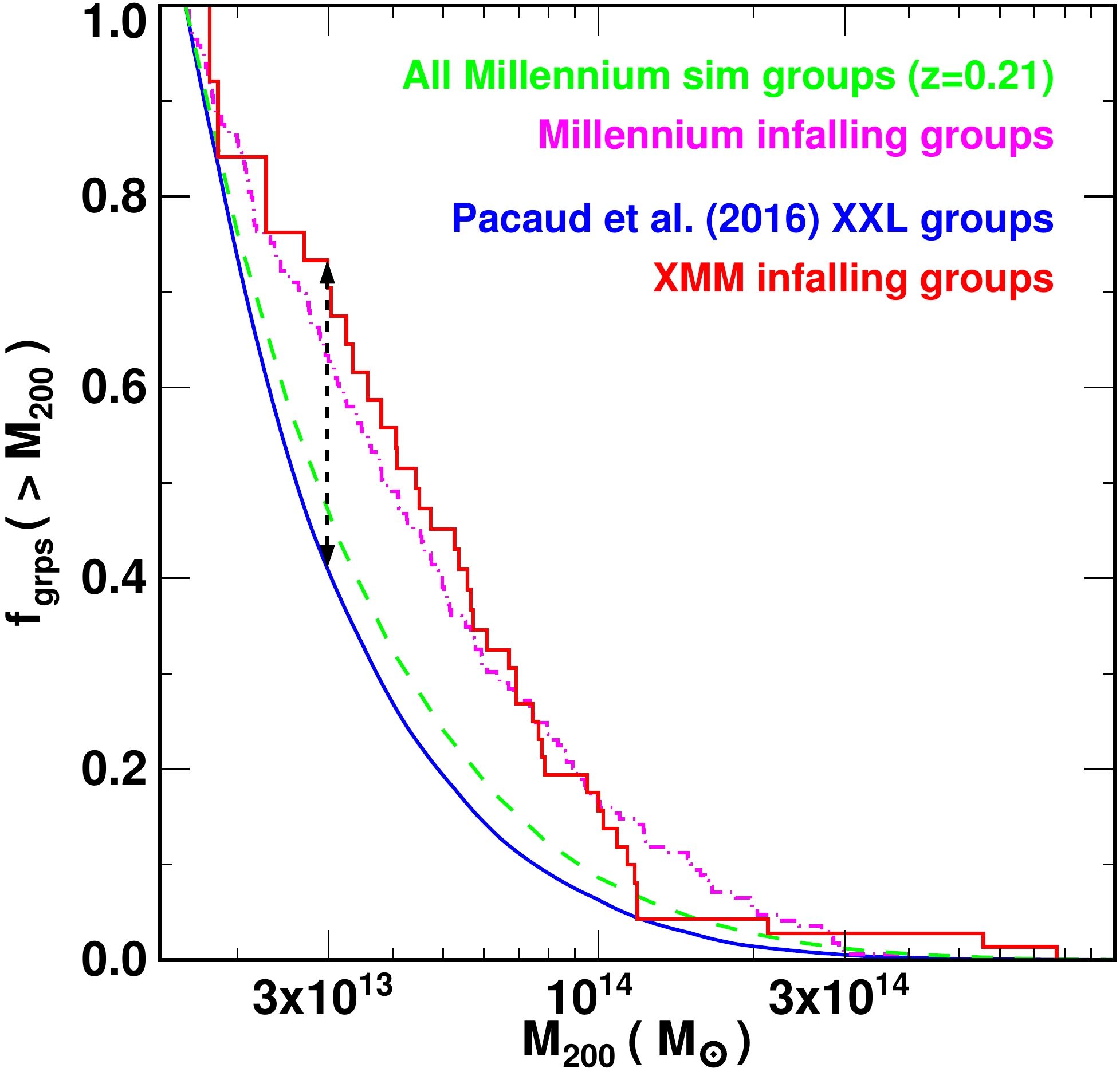}}
\caption{Cumulative mass distribution $f({>}M)$ of the 39 {\em XMM}-detected galaxy
  groups ({\em red line}), X-ray groups from the XXL survey
  \citep[{\em blue curve}]{pacaud}, 
infalling galaxy groups around 75 massive clusters in the Millennium simulation ({\em dot-dashed magenta line}), and all
  $z{=}0.21$ DM halos with $M_{200}{\ge}10^{13.2\,}{\rm M}_{\odot}$ in the
  Millennium simulation ({\em green curve}). The vertical dashed line
  indicates the maximal differences between the observed cumulative mass distributions.}
\label{ks_test}
\end{figure} 

The apparent differences in the shape of the mass function of X-ray groups according to
large-scale environment is further elucidated in Figure~\ref{ks_test},
which shows the cumulative mass fractions of  X-ray groups from the XXL survey ({\em blue curve}) and
the 39 infalling X-ray groups ({\em red line}), as well as the
corresponding DM halo populations from the Millennium simulation. 
Each curve presents the fraction of M$_{200}{>}10^{13.2\,}{\rm
  M}_{\odot}$ groups (DM halos) that are also above a given M$_{200}$
mass, as a function of M$_{200}$. This confirms that the mass function
of the X-ray groups found in the vicinity of massive clusters is
systematically top-heavy with respect to that of the general population of X-ray
groups at these redshifts from the XXL survey. A Kolmogorov-Smirnov
test finds that the probability that both mass functions are drawn
from the same distribution to be just 0.0006 ($D_{KS}{=}0.323$;
maximal distance between curves is shown by the vertical black
line). This corresponds to a 3.5$\sigma$ result.

The systematic bias towards a top-heavy MF observed for X-ray groups
around massive clusters replicates that seen in the Millennium
simulation \citep{faltenbacher,haines15}. The cumulative mass function of DM halos in the vicinity ($r_{proj}{<}1.3\,r_{200}$) of a
massive cluster at $z{=}0.21$ ({\em magenta dot-dashed curve}) is found
to be consistent with our observed MF for infalling X-ray groups, and
top-heavy with respect to the MF of DM halos averaged across the full
Millennium simulation at the same snapshot ({\em green dashed curve}). 
The large-scale overdensity centred
on the massive cluster biases the halo mass function in its vicinity,
increasing the relative contribution of higher mass halos
at the expense of lower mass
systems. This biasing has the effect of increasing the importance of
accreting ${\sim}10^{14}\,{\rm M}_{\odot}$ systems to the mass growth
of rich clusters with respect to simple predictions based on the
cosmic halo mass function. 


\subsection{The total mass contained within infalling groups}

The total mass of the 39 infalling groups detected by {\em XMM} is
$3.77{\times}10^{15}\,{\rm M}_{\odot}$, which after correcting for
incompleteness comes to $5.13{\times}10^{15}\,{\rm M}_{\odot}$ in
systems above $10^{13.2}\,{\rm M}_{\odot}$, or
$2.23{\times}10^{14\,}{\rm M}_{\odot}$ for each of the 23 clusters in
our sample. This corresponds to $19.5{\pm}5.1$\% of the mean mass of
the 23 primary clusters ($\langle
M_{200}\rangle{=}11.47{\times}10^{14\,}{\rm M}_{\odot}$), where the
uncertainty is estimated by bootstrap resampling to account for 
the significant cluster-to-cluster scatter. 

The two most massive X-ray ``groups'' in
Fig.~\ref{massfunction} are Abell 1758S and Abell 115S, both of which
are the lesser component of well-known double clusters undergoing
major mergers \citep{david,gutierrez,okabe08}. Although nominally clusters, we include these among our
infalling X-ray ``groups'' as they will be accreted and subsumed by
the primary cluster (A1758N, A115N) in the same way. Even so, these
two clusters only contribute $1.33{\times}10^{15\,}{\rm M}_{\odot}$
between them, representing 26\% of the total mass within our infalling X-ray group
sample. Excluding them does not dramatically change our estimate of
the amount of mass being accreted onto clusters in the form of
groups. 


\subsection{The group-cluster mass ratio distribution}

\begin{figure}
\centerline{\includegraphics[width=84mm]{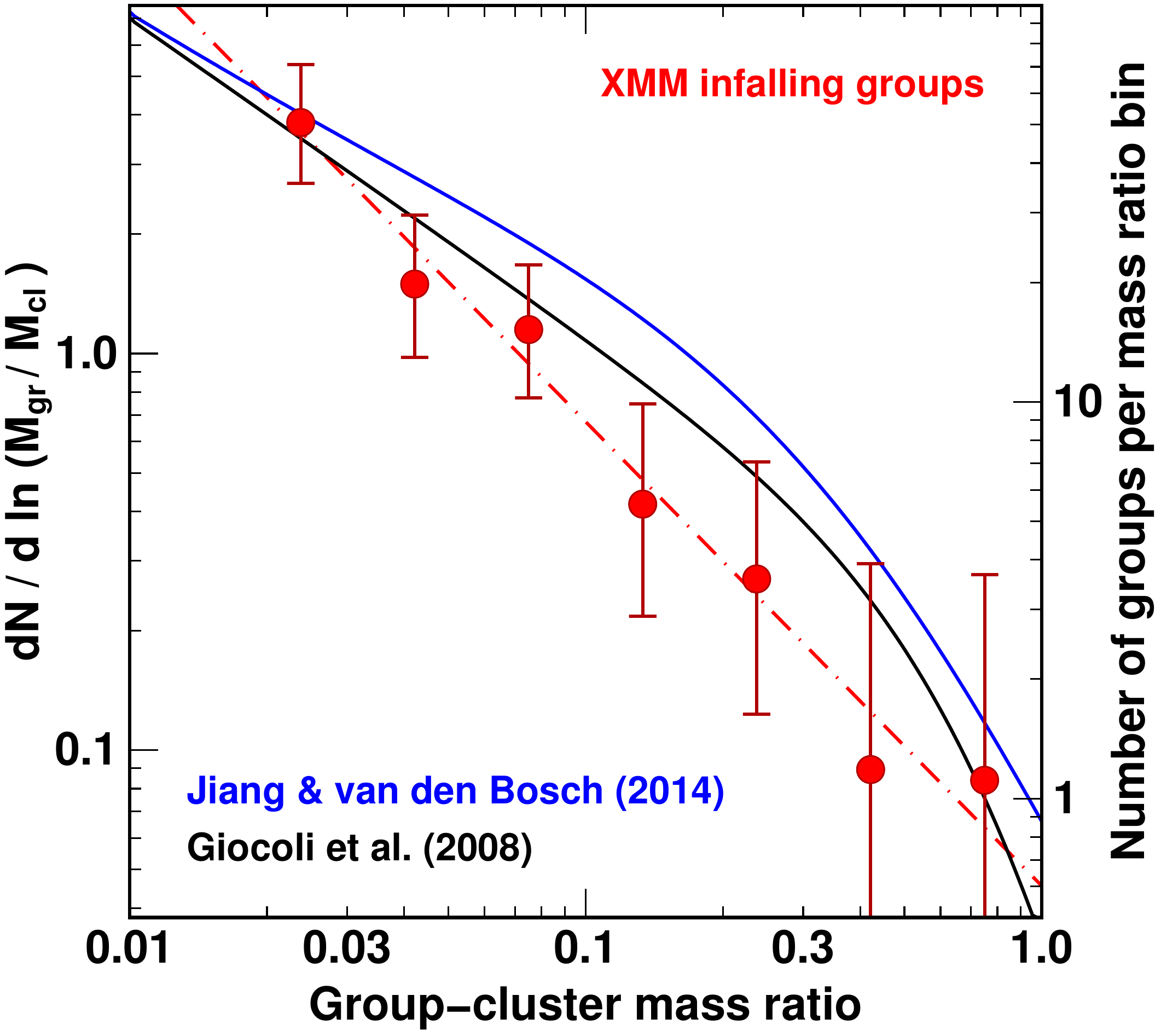}}
\caption{The distribution of the mass ratios between {\em XMM}-detected galaxy
  groups and the primary clusters, after correcting for incompleteness
  ({\em red solid symbols}). The best-fit power law is shown by the red
  dot-dashed line. The black and blue curves show the predicted {\em unevolved} subhalo mass
  functions of \citet{giocoli08} and \citet{jiang} respectively. 
}
\label{massratios}
\end{figure}

Figure~\ref{massratios} shows the distribution of $M_{200}$ mass ratios between
the infalling galaxy groups and the primary clusters that they are
associated with ($M_{gr}/M_{cl}$; {\em red points}), using the {\em XMM}-based
M$_{200,X}$ cluster masses from Table~\ref{clusters}. The mass-ratio
distribution appears approximately consistent with a power law
$dN/d\,{\rm ln}\,(M_{gr}/M_{cl}){\propto}(M_{gr}/M_{cl})^{-\alpha}$ over
the range 0.02--1.0 in group-cluster mass ratio, with
$\alpha{=}1.17^{+0.28}_{-0.34}$ ({\em red dot-dashed line)}.
This is in excellent agreement with
the best-fit power-law index of $1.09^{+0.42}_{-0.32}$ obtained by
\citet{okabe14} for the sub-halo mass function through a weak lensing
analysis of sub-halos in the Coma cluster, and indices
${\sim}0$.9--1.0 predicted by analytical models \citep{taylor05} and numerical simulations.

The black solid curve shows the {\em unevolved} subhalo mass function of
\citet{giocoli08}:
\begin{equation}
\frac{dN}{d\ln (m_{gr}/M_{cl,0})} = N_{0} x^{-\alpha} \exp
(-6.283x^{3}),\, x{=}\left| \frac{m_{gr}}{\alpha M_{cl,0}} \right|
\end{equation}
where $m_{gr}$ is the mass of the progenitor group halo at the time
of accretion, $M_{cl,0}$ is the present day mass of the descendent
cluster, $\alpha{=}0.8$ and $N_{0}{=}0.21$. Here we assume that
$M_{cl,0}{=}(M_{cl}{+}M_{gp})$, i.e. the group's mass has been
subsumed by the cluster by the present day. 
\citet{jiang} refined the fitting function of \citet{giocoli08}, adding
in an extra power-law term to better model the unevolved
subhalo mass function of halos within the Millennium simulation ({\em
  blue curve}).

The form and steepness of the mass-ratio distribution of infalling {\em XMM}
groups reproduces well the unevolved subhalo mass functions of
both \citet{giocoli08} and \citet{jiang}, the main difference being a
systematic shortfall at most mass ratios. This is unsurprising, as we
are only detecting the groups which are being accreted into the
clusters at late epochs, while the functions of \citet{giocoli08} and
\citet{jiang} include the contributions of subhalos accreted at all
redshifts, a significant fraction of which will have long been
stripped of their X-ray emitting gas halos.

\subsection{Offsets between the BGG and the X-ray centroid of the
  infalling groups}

In isolated, undisturbed galaxy groups, the central group galaxy (BGG)
and the centroid of the X-ray emission should be coincident. When a
group falls into a galaxy cluster, the X-ray emitting gas of the group is
incrementally stripped by the ram pressure exerted by the cluster's own X-ray
emitting ICM. This causes the group's X-ray gas to drag and thus lag
behind the member galaxies, which as effectively collisionless
particles. For the 32 infalling X-ray groups for which a 
clear BGG could be identified, we find a median separation of 65\,kpc between the BGG and
the X-ray centroid, with 68\% of the separations in the range
28--143\,kpc. These separations are consistent with those seen for
infalling groups in the BAHAMAS simulation \citep{mccarthy17}.
The X-ray centroid is further away from the cluster
centre than the BGG for 19/32 systems.

\section{Discussion}

A fundamental prediction of the $\Lambda$CDM model is that galaxy clusters, as
the most massive collapsed halos in the universe, form latest,
doubling their mass since $z{\sim}0.5$ \citep[e.g.][]{van14}. As structure formation occurs
hierarchically, much of this late mass growth must be achieved through the
accretion of poorer clusters and group-mass systems, and so the outer
regions of clusters must be replete with infalling group-mass
systems. The key objective of our {\em XMM} survey of 23 massive
clusters is to perform a simple empirical verification of the ongoing
assembly of massive clusters through the accretion of groups, as
predicted by $\Lambda$CDM, and to estimate the contribution of these
infalling groups to the mass growth rate of the primary clusters.


\subsection{The mass assembly history of clusters}

The average rates at which clusters assemble their mass through
mergers and accretion as function of redshift have been investigated
for a range of cosmologies using a combination of N-body simulations
and Monte-Carlo realisations based on the extended Press-Schechter (EPS)
framework.

\begin{figure}
\centerline{\includegraphics[width=84mm]{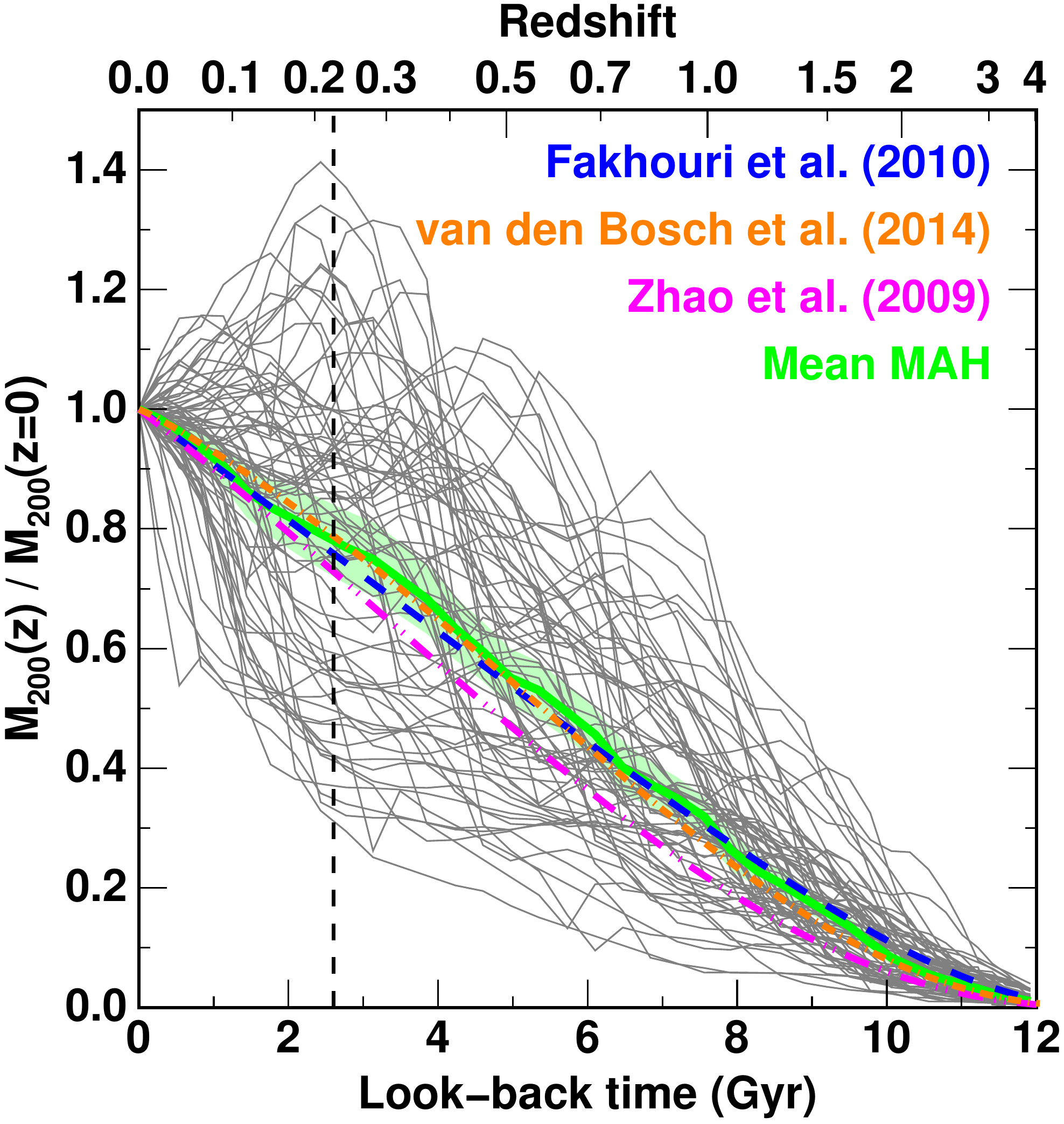}}
\caption{The mass accretion histories (MAHs) of the 75 most massive DM halos in
the Millennium simulation (grey curves). The solid green
curve indicates the mean MAH of these 75 halos, while the green shaded
region indicates the expected $1\sigma$ range of MAHs when averaging
over 23 clusters selected at random from the full sample. 
The dashed blue, dot-dashed orange and magenta curves indicate the mass growth for
halos of present day mass M$_{200}{=}10^{15}h^{-1}\,{\rm M}_{\odot}$ predicted by
Eq.~\ref{massgrowthrate} \citep{fakhouri10}, \citet{van14} and
\citet{zhao09} respectively. The vertical black dashed line indicates
the mean redshift ($\overline{z}{=}0.223$) of the LoCuSS primary clusters and {\em XMM}
infalling groups.
}
\label{mill_mahs}
\end{figure}

\citet{mcbride} investigated the mass accretion histories (MAHs) of
DM halos from the Millennium simulation, finding that a
two-parameter function of the form
\begin{equation}
M(z)= M_{0} (1+z)^{\beta} \exp(-\gamma z)
\end{equation}
was versatile enough to accurately capture the main features of most
MAHs in the simulation. They were also able to obtain a good fit to
the mean mass growth rates of halos as a function of halo mass and
redshift, by differentiating the above equation. \citet{correa} have
shown using EPS theory and the redshift dependence of the linear growth factor
$D(z)$ that the mass growth of halos is well
described by an exponential growth at high redshifts, while at low redshifts when dark
energy dominates, the growth of density perturbations is slowed by the
accelerated expansion of the Universe, necessitating an additional
power-law term. 

\begin{figure*}
\centerline{\includegraphics[width=120mm]{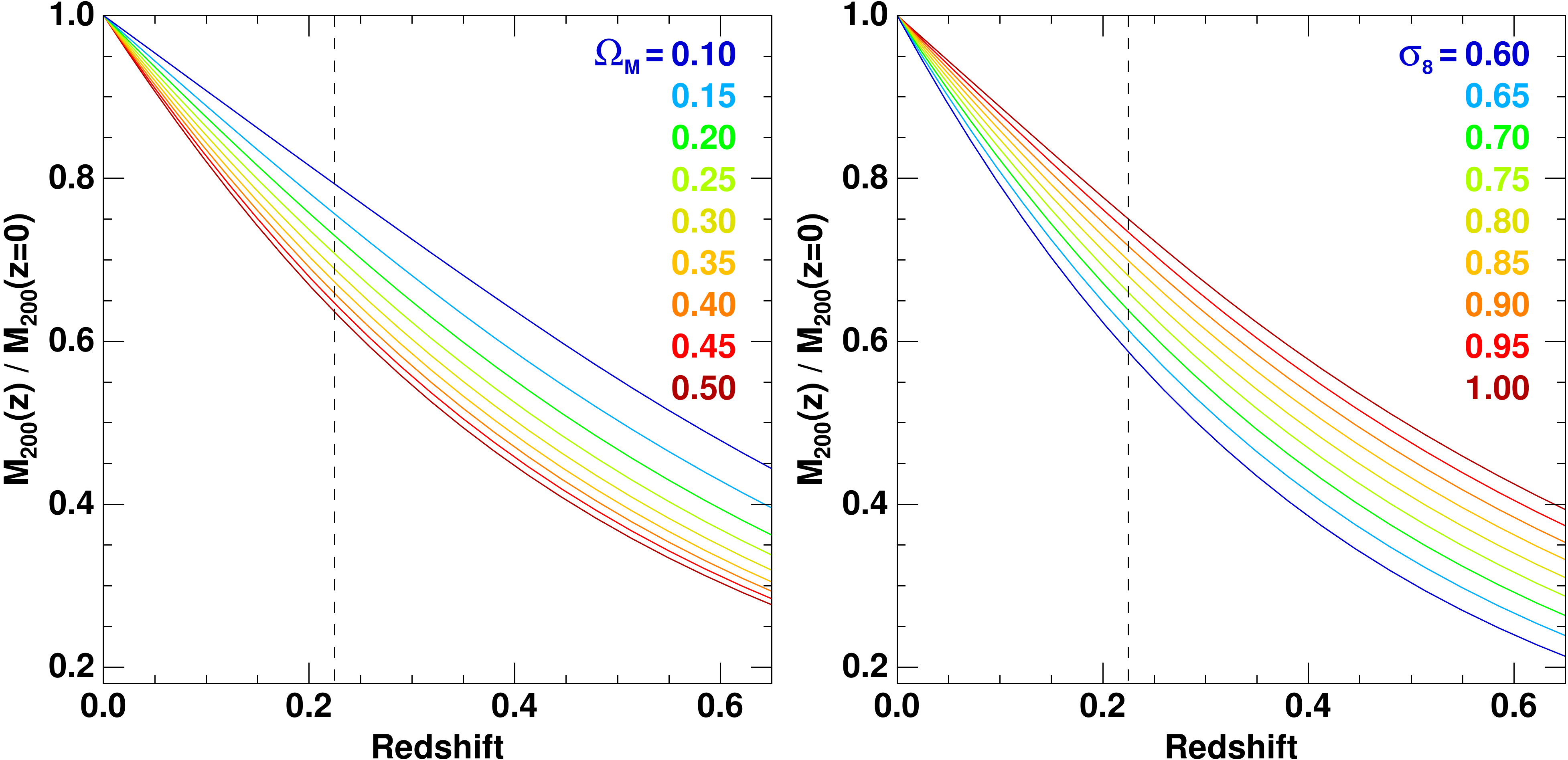}}
\caption{Average MAHs, $\langle M_{200}(z)/M_{200}(z=0)\rangle$, for
  halos of final mass M$_{0}{=}10^{15}h^{-1}\,{\rm M}_{\odot}$ in
  different, flat $\Lambda$CDM cosmologies, based on the empirical
  models of \citet{zhao09}. The left panel shows the
  effect of varying $\Omega_{M}$ from 0.1 to 0.5 in steps of 0.05, while the right panel shows the
  effect of varying $\sigma_{8}$ from 0.6 to 1.0 in steps of 0.05. All
  of the other cosmological parameters are kept fixed at the values
  from the Planck 2015 model. The vertical dashed line indicates the
  mean redshift ($\bar{z}{=}0.223$) of our sample of 23 clusters.} 
\label{cosmoMAH}
\end{figure*}

 Figure~\ref{mill_mahs} shows the mass accretion histories of the 75
most massive DM halos in the Millennium simulation (MS; {\em grey
  curves}). The main branch of the merger tree of each cluster halo is
determined by identifying the most massive progenitor of the
descendent cluster halo in the previous snapshot. The most notable
feature is the large cluster-to-cluster scatter among the MAHs, with
some clusters assembling more than half of their mass in the last
2\,Gyr, while others were largely in place by $z{\sim}0.5$ and some
are appearing to be losing mass at late epochs. 
These mass loss events occur during major mergers, which produce first a dramatic increase in
mass and a corresponding rapid increase in the velocity dispersion,
followed by a slower phase of mass loss as a significant amount of
mass from the secondary halo rebounds out of the primary halo and
orbits beyond its $r_{200}$ radius for 2--4\,Gyr
\citep{behroozi}, analogously to ``back-splash'' galaxies. 

 While the individual cluster MAHs show significant stochastic
variation, the mean MAH ({\em green curve}) shows a steady monotonic
increase in mass over the last 10\,Gyr. The large scatter in individual MAHs implies the need
to average over many clusters to derive useful constraints on the
cosmic growth of mass within clusters, and the green shaded region
indicates the $1\sigma$ range in {\em averaged} MAHs produced by
combining the growth rates of 23 clusters, selected at random from the
full sample. The total $M_{200}$ masses of a random sample of 23 MS clusters should
have $77.6{\pm}5.9$\% of their present day $M_{200}$ mass at
$z{=}0.223$, the mean redshift of our primary clusters ({\em vertical
  dashed line}). In other words, they should grow by a further
$29{\pm}10$\% between $z{=}0.223$ and the present day. 

Using the higher-resolution Millennium II simulation,
\citet{fakhouri10} obtained updated analytic fits to the mean mass growth rates
of halos of mass $M$ at redshift $z$ of:
\begin{multline}
\langle \dot{M} \rangle_{\rm mean} =  46.1\, {\rm M}_{\sun}\, {\rm yr}^{-1} \left(
  \frac{M}{10^{12}{\rm M}_{\sun}} \right)^{1.1} \\
  {\times} (1+1.11z) \sqrt{\Omega_{m}(1+z)^{3} + \Omega_{\Lambda}}.
\label{massgrowthrate}
\end{multline}
The resulting mass accretion history for a cluster halo of present day
mass $M_{200}{=}10^{15\,}{\rm M}_{\odot}$ ({\em dashed blue 
  curve}) is consistent with the mean MAH of the 75 most massive
MS clusters. 
\citet{zhao09} derived a
universal empirical model for the MAHs of DM halos, by analysis of
numerous N-body simulations of a wide variety of cosmological models,
which predicts a somewhat later mass assembly for
$10^{15}h^{-1}\,M_{\odot}$ halos ({\em dot-dashed magenta curve}) in a
Millennium cosmology. 
\citet{van14} used EPS merger trees calibrated with the Bolshoi N-body 
simulation to derive average MAHs for halos of a given mass in
any $\Lambda$CDM cosmology ({\em dot-dashed orange curve}).
 
These models all predict similar MAHs for $10^{15}h^{-1}\,{\rm M}_{\odot}$
clusters, whereby they have grown at a virtually
constant rate since $z{\sim}1$ (Fig.~\ref{cosmoMAH}). 
The MAHs of \citet{van14}, \citet{fakhouri10} and \citet{zhao09} respectively predict 
mass increases of 27\%, 32\% and 37\% between $z{=}0.223$ and the
present day. These correspond to mass growth rates of 13-17\% per Gyr for clusters at $z{\sim}0.2$. 


\subsubsection{Dependence on cosmological parameters}

The curves shown in Fig.~\ref{mill_mahs} demonstrate well the expected
cluster-to-cluster scatter among their MAHs. Strictly speaking they
are only valid for the exact cosmological model
used in the simulation. The parameters were set in the MS ($\Omega_{m}{=0.25}, \Omega_{\Lambda}{=}0.75, \sigma_{8}{=}0.90,
n_{s}{=}1.0, h{=}0.73$) to be close to those obtained from
WMAP1 \citep{spergel}, but are now somewhat divergent from the current best estimates obtained from
analysis of the full-mission Planck CMB data \citep[$\Omega_{m}{=}0.308{\pm}0.012$,
$\sigma_{8}{=}0.8149{\pm}0.0093$, $n_{s}{=}0.9677{\pm}0.0060$,
$H_{0}{=}67.81{\pm}0.92$;][]{planck2015}.

The universal models of \citet{zhao09} and the EPS-based models of
\citet{van14}, allow us to investigate the dependence of the average MAHs of
cluster-mass halos on the main cosmological parameters, and recalibrate the
results onto the Planck 2015 cosmological model.
Figure~\ref{cosmoMAH} shows the effect of varying $\Omega_{m}$ and $\sigma_{8}$ on the average MAHs of DM halos with present
day masses $M_{200}{=}10^{15}\,{\rm M}_{\odot}$ in flat $\Lambda$CDM
cosmologies, while keeping the remaining parameters fixed to the
Planck 2015 values, using the \citet{zhao09}
code with an \citet{eisenstein} power spectrum. These confirm that both $\Omega_{m}$ or $\sigma_{8}$ have a
significant affect on the MAHs of cluster-mass halos, with the rate of
growth at late epochs rising with increasing $\Omega_{m}$ and decreasing $\sigma_{8}$.
In contrast, varying the tilt of the primordial power-law spectrum
$n_{s}$ has negligible influence on the MAHs of cluster-mass halos.

Increasing $\Omega_{m}$ from 0.25 (MS) to 0.308 (Planck 2015) results in a
corresponding increase in the rate of growth of clusters from
$z{=}0.223$ to the present day of 7.4\%, while decreasing $\sigma_{8}$
from 0.90 to 0.8149 increases the growth rate by an additional
10.9\%. The combined changes in moving from the MS 
cosmological model to that of Planck 2015 increases the late-time
growth rate by 18.3\%, resulting in mass growth rates for $z{\sim}0.2$
clusters of 15--20\% per Gyr.


\subsection{The contribution of X-ray groups to the mass growth rate of clusters}

Our finding of 39 {\em XMM}-detected galaxy groups in the infall
regions of 23 massive clusters, corresponding to an average mass of
$2.23{\times}10^{14}\,{\rm M}_{\odot}$ per cluster, or 19\% of the
mean M$_{200}$ mass of the primary cluster, suggests that these galaxy
groups can explain a significant fraction of the mass growth of
galaxy clusters between $z{\sim}0.2$ and the present day. 

 The comparison sample of group-mass halos from the Millennium simulation allow us examine the
likely fates of the 39 {\em XMM}-detected galaxy groups, by following
the orbits from the $z{=}0.21$ snapshot where they are identified to
the present day. We find that two-thirds (67\%) of the simulated groups located
within the {\em XMM} fields and recovered by the
wavelet-reconstruction algorithm will be accreted into the cluster
(pass within $r_{200}$) by the present day. In contrast, 22\% of the
detected {\em XMM} groups are expected to be physically more than
4\,$r_{200}$ from the cluster at the time of observation and despite
being on their first infall, remain outside $r_{200}$ at $z{=}0$.
This is balanced by the 24\% of those simulated infalling 
M$_{200}{>}10^{13}\,{\rm M}_{\odot}$ halos accreted into the clusters
between $z{=}0.21$ and $z{=}0$ being outside the {\em XMM} field of view at $z{=}0.21$ and
therefore missed from our sample. This can be understood given that
those halos only accreted into the clusters in the last snapshot ($z{<}0.02$) were on average at
$2.89{\pm}0.70\,r_{200}$ at $z{=}0.21$. 

Taking the above correction factors into account, we estimate that
clusters increase their mass by $16.2{\pm}4.2$\% between $z{=}0.223$ and the
present day (or 6\% per Gyr) due to the accretion of groups more massive than
$10^{13.2}\,{\rm M}_{\odot}$. This confirms that X-ray groups are
contributing significantly to the mass growth rate of clusters. 
However, this estimate for the mass accreted in the form of groups is
only half that predicted for the overall mass growth of massive
clusters over the same period (32--44\%), as described in the previous section.


\subsection{Accounting for the rest of the mass accreted by clusters
  and estimating their growth rates}

Our empirical estimate that clusters are able to increase their masses by
${\sim}1$6\% between $z{=}0.223$ and the present day through the
accretion of $M_{200}{>}10^{13.2\,}{\rm M}_{\odot}$ X-ray groups is
not sufficient to fully explain the mass growth rate of clusters. 
Thus, either the growth rate of massive clusters is much lower than
that predicted by cosmological simulations, or there are other major
contributions to the mass accretion rate of clusters from less massive
DM halos (e.g. those hosting individual galaxies) or matter that
is not bound within any DM halo. 

Within the extended Press-Schechter (EPS) formulism \citep{press,bond,bower,lacey}, all of the growth of
dark matter halos comes from mergers by construction. However, 
using merger trees constructed from both Millennium
simulations, and taking care to accurately account for halo
fragmentation, \citet{genel10} find that all resolved mergers, down to
mass ratios of $10^{-5}$ between them contribute only ${\approx}$60\%
of total halo mass growth, regardless of halo mass and redshift. 
Major mergers with ratios above 1:3 (1:10)
contribute just 20\% (30\%). Instead they indicate that 40\% of the
mass in halos (up to and including cluster-mass halos) comes from genuinely smooth accretion of dark matter
that was never bound in smaller halos. While there is some freedom of
how merger trees are constructed, \citet{genel10} verified this result
by following the individual dark matter particles within two 
cosmological simulations and labelling each one
that had belonged to an identified bound structure at any point in its past,
prior to its accretion into the primary halo. 

Our estimate that the accretion of $M_{200}{>}10^{13.2\,}{\rm
  M}_{\odot}$ X-ray groups are sufficient to account for roughly
35--50\% of the predicted mass growth rate of ${\sim}10^{15\,}{\rm
  M}_{\odot}$ clusters, appears consistent with the findings of
\citet{genel10}.  
This assumes that our $L_{X}$-based group mass estimates are
unbiased relative to the true masses. Given their proximity to the
primary cluster, some of these groups could be affected by ram-pressure
stripping which progressively removes the X-ray emitting gas (see
$\S$~\ref{next}). 
As a sanity check of our mass estimates, we split the groups into
three mass bins, and for those groups with 4 or more confirmed
members, measure the distribution of the LOS velocity offsets relative
to the group's mean redshift. The resultant velocity dispersions are
257\,km/s for the 12 groups with $\log M_{200}{<}13.75$, 328\,km/s for
the 14 groups with $13.75{\le}\log M_{200}{<}14.05$ and 369\,km/s for
the 7 groups with $\log M_{200}{>}14.05$. These values are consistent with
the $M_{200}-\sigma$ trend of SDSS groups \citep{yang07} and the
$L_{X}-\sigma$ relations of \citet{zhang11} and \citet{clerc}. The LOS
velocity distributions of each stacked group sample are consistent
with being a Gaussian function. 
Our sample may however miss the mass contribution from 
groups which have been recently accreted but are now orbiting back out
beyond $r_{200}$. Their dark matter halo may still be largely
intact, but the X-ray emitting gas has been sufficiently stripped as
to be undetected. The LOS velocity offsets of such groups are likely
to be rather low as they approach apocenter. 

\citet{deboni} have suggested that it is possible to estimate the
overall mass accretion rate of clusters from their mass profiles
beyond the virial radius. The
agregate radial velocity of dark matter within a radial shell reaches
a minimum at 2--3\,$r_{200}$, that is beyond the splashback radius,
and so most closely represents the infall of new material onto the
cluster. By measuring the mass profile of the cluster over
2--3\,$r_{200}$ using the caustic method of \citet{diaferio}, and
assuming the infall velocity based on spherical collapse model, they
are able to approximately reproduce the mass accretion rates of
clusters within simulations. Given the current availability of dense redshift
surveys of clusters galaxies beyond 2\,$r_{200}$ for many rich
clusters \citep[e.g.][]{rines}, estimates of their typical mass
accretion rates should be feasible.


\subsection{The mass function of infalling X-ray groups}

The variation of the DM halo mass function with large-scale density
such that the MF appears top-heavy in overdense regions can be readily
understood from a theoretical perspective.
Collapsed DM halos are biased tracers of mass. This implies that the
abundance of DM halos in overdense and underdense regions are not
expected to simply differ by a factor which reflects the change in
large-scale matter density. Instead:
\begin{equation}
n(M|\delta)\approx[1+b(M,z)\delta]\,n(M)
\end{equation}
where $n(M|\delta)$ is the abundance of DM halos of mass $M$ in a
region of overdensity $\delta$,  $n(M)$ is the cosmically averaged
abundance, and $b(M,z)$ is the mass-dependent bias
parameter of halos at redshift $z$ \citep{mo,sheth,abbas}. As the bias $b(M)$ typically
increases monotonically with mass \citep{tinker10}, this acts to increase the ratio of
high-mass DM halos to low-mass halos in overdense regions, relative to
less dense regions. Thus the mass function in overdense regions should
be top-heavy.  The effects of large-scale density on the halo mass
function were examined by \citet{faltenbacher} using the Millennium
simulation, confirming that the fraction of matter within group-mass
halos ($M_{200}{\ga}10^{13.5\,}{\rm M}_{\odot}$) increases
significantly with large-scale density, and the halo mass function
becomes increasingly top heavy \citep[see also][]{lemson}.

\citet{chon13} found that clusters within
superclusters were systematically more X-ray luminous than clusters
outside superclusters. Assuming that this overabundance of X-ray
luminous clusters represents an excess of massive clusters within
superclusters, provides observational support for the theoretical expectation that the mass-function of clusters in
overdense regions (superclusters) is top heavy. Similarly, this mass
bias was also observed for galaxy groups in the vicinity of clusters
in the 2dF Galaxy Redshift Suryve \citep{ragone}.


\subsection{Impact for galaxy evolution}
\label{impact}

These X-ray groups are not only contributing a large proportion of the 
the dark matter required for cluster mass growth, but also host a  
significant fraction of the galaxies that arrive onto the clusters at
late epochs. \citet{mcgee} estimate that ${\sim}5$0\% of cluster
galaxies accreted since $z{=}0.5$ arrived onto the cluster as member
of an infalling group with M$_{200}{>}10^{13\,}{\rm M}_{\odot}$.
Galaxy groups have been shown to have a major impact on the evolution
of their member galaxies, suppressing star formation activity through
the interaction of the galaxy with the intra-group medium
(ram-pressure stripping or starvation), or transforming their morphologies through low-velocity encounters and mergers with other group
members. The fraction of star-forming galaxies among group members is
lower than that seen in the field \citep[at fixed stellar mass and redshift;][]{haines07,ziparo}, and declines with increasing group
mass and proximity to the group centre \citep{weinmann,woo}. 

Thus, many galaxies are arriving onto clusters having already been
transformed from star-forming spirals into passive early-types within
groups, a mechanism known as pre-processing \citep{zabludoff96,zabludoff98,dressler,just,jaffe}. This can contribute
signficantly to the cluster population of passive early-types, but
also explain the short-fall of star-forming galaxies at large
cluster-centric radii ($\ga\,$2--3\,$r_{200}$) where no galaxies should have
previously encountered the cluster \citep{chung,haines15}. This should be
exacerbated by the top-heavy mass function of these infalling groups,
meaning that galaxies are more likely to be in massive X-ray luminous groups than the cosmic average. 
We will examine the impact of pre-processing on the galaxies within these infalling X-ray groups in
\citet{bianconi}. 


\subsection{The next steps}
\label{next}

This work presents a first attempt to quantify the numbers and
demographics of X-ray groups in the immediate vicinity of a
statistical sample of massive clusters, and derive empirical constraints on
the rates at which clusters are growing through the accretion of
group-mass systems. By identifying groups through their extended X-ray
emission, we can confidently associate them to massive virialized DM
halos. As we only have the X-ray luminosities of these groups, we have
had to make certain simplifying assumptions to estimate their masses. 
In particular, by using the $M_{200}-L_{X}$ scaling relation of
\citet{leauthaud}, we are assuming that the X-ray emitting gas content
of these infalling groups remains bound within the host DM halo, and maintains
the same density and temperature structures as isolated
field groups (such as those from the COSMOS survey).
At the same time, we expect that as these groups are accreted into the
cluster, passing through the increasingly dense ICM, their X-ray
emitting hot gas halos are progressively ram-pressure stripped
\citep{gunn,poole,mccarthy}. 
The plasma physics of this process however very
complex, with magnetic fields, turbulence, viscosity, KH instabilities and
conduction all likely to play a role in determining when and how rapidly X-ray
emitting gas is stripped from the group, the appearance of the
extended tail of high-density stripped gas and how long this wake
can survive before mixing with the ambient ICM \citep{roediger,roediger2}. 

Examples of this gas strippping have been seen in recent X-ray
observations of groups infalling into Abell\,85, Abell\,2142,
Abell\,4067, ZwCl\,8338 and Abell 780 \citep{ichinohe,eckert,chon,schellenberger,degrandi}.
These group-cluster mergers also leave shock fronts, spiral features
indicative of gas sloshing and increased gas clumping in the cluster
outskirts \citep{reiprich}. On the other hand, during the
group-cluster mergers, the X-ray luminosity may be briefly boosted as
the group makes its pericenter passage \citep{ricker}, biasing the
resulting group mass function \citep{randall}. The hydrodynamical
simulations of group-cluster mergers by \citet{poole} show how as a
group approaches pericenter, the gas on its leading edge is heated and compressed,
temporarily boosting its X-ray luminosity, before being steadily stripped, leaving
an extended trail of cool, low-entropy gas behind it, similar to those
seen by \citet{eckert} and \citet{degrandi}. In particular,
\citet{eckert} estimate that ${>}9$0\% of the gas mass from the group
falling into Abell 2142 has been stripped to form an 800\,kpc long tail. 
These detailed observations of individal group-cluster mergers are
providing fundamental insights and constraints on the ICM plasma physics
involved, revealing that the thermalization and mixing of the stripped
group gas must be slow and inefficient \citep{eckert17}, and should
lead to a better understanding of when and how the X-ray emitting gas
is stripped from infalling groups. 

The centres of an isolated group's galaxy population, X-ray gas and DM halo
should be coincident. As a group falls into a cluster, the ram
pressure acts as a drag on the X-ray gas, causing it to lag behind the
member galaxies and the dark matter, both of which are effectively
collisionless. Hubble Space Telescope observations of these groups
would allow their dark matter distributions to be
constrained, to both confirm the overall masses of these groups and
evidence of this lag in the hot gas component relative to the dark
matter. Such observations can also test the self-interaction cross
section of dark matter, which would create a drag on the dark matter
within the infalling groups, displacing the DM distribution relative
to the group galaxies \citep{harvey,harvey15}.

This {\em XMM} survey of infalling groups should also be a precursor to the much larger 
samples that should be obtained in the near future with eROSITA \citep{merloni}. 
eROSITA will aim to perform a deep X-ray survey of the entire sky, with a combination 
of resolution and sensitivity good for studying the galaxy groups
around $z<0.1$ clusters. It will lead to large statistical improvements primarily 
on lower mass clusters, compared to the current sample, as the local 
volume is small. Also, in a combination with spectroscopic follow-up on 
4MOST/VISTA it will deliver better statistics on larger separations from 
the cluster centre.

\section{Summary}

We present an {\em XMM-Newton} survey to search for X-ray groups in
the infall regions of 23 massive galaxy clusters at
$0.15{\le}z{<}0.30$ from the Local Cluster Substructure Survey
(LoCuSS). All these clusters have excellent ancillary data including
extensive spectroscopic coverage of cluster galaxies through the
Arizona Cluster Redshift Survey (ACReS) and deep wide-field optical
imaging from Subaru/Suprime-Cam, enabling us to identify the member
galaxies associated with the X-ray emission of each group, and
determine its redshift.

We identify 39 X-ray groups across the 23 {\em XMM} fields that have been spectroscopically confirmed
to lie at the cluster redshift (and hence are likely falling into the
primary cluster), above a signal-to-noise limit of 3. These groups
all have at least one spectroscopic member, and a median of nine
members. 
These infalling groups have $M_{200}$ masses in the range
$2{\times}10^{13}{-}7{\times}10^{14\,}{\rm M}_{\odot}$, based on
estimates derived from their X-ray luminosities. The key results from
a statistical analysis of these groups are:
\begin{itemize}
\item The 39 infalling X-ray groups lie at \mbox{0.28--1.35}\,$r_{200}$
  and are much more concentrated towards the cluster than the other 52
  groups in the same fields (at the $6\sigma$ level). The distribution
  of the LOS velocity offsets of the infalling groups relative to the
  primary clusters is non-Gaussian, consistent with them being an
  infalling population.
\item The comoving number density of the infalling X-ray groups is
  ${\sim}25{\times}$ higher than that seen in field regions. 
\item The mass function of the infalling X-ray groups is top-heavy
  with respect to that seen for isolated groups in the XXL survey at the 3.5$\sigma$
  level. This is consistent with expectations
  of collapsed DM halos being biased tracers of the
  underlying large-scale density field.
\item The average mass per cluster contained within these infalling
  X-ray groups is $2.2{\times}10^{14\,}{\rm M}_{\odot}$, or
  $19{\pm}5$\% of the mass of the primary cluster.
\item We estimate that ${\sim}10^{15\,}{\rm M}{\odot}$ clusters
  increase their masses by $16{\pm}4$\% between $z{=}0.223$ and the
  present day due to the accretion of X-ray groups with
  $M_{200}{\ga}10^{13.2\,}{\rm M}_{\odot}$. This represents 35--50\%
  of the expected mass growth of these clusters at these late
  epochs. The rest of the mass growth is likely to occur through
  the smooth accretion of dark matter not bound within DM halos.
\end{itemize}

This work represents the first attempt to statistically establish the
frequency and demographics of X-ray groups in the infall regions of
a representative sample of massive clusters, estimate their contribution to the mass growth of
clusters and mass function. It complements ongoing detailed X-ray studies
examining the astrophysical processes acting on group-mass systems as
they are accreted into massive clusters \citep[e.g.][]{eckert,degrandi}. 

\section*{Acknowledgments}

CPH acknowledges financial support from PRIN INAF 2014 and CONICYT
Anillo project ACT-1122. GPS, MB, and FZ acknowledge support from the Science 
and Technology Facilities Council grant number ST/N000633/1.

\bsp

\label{lastpage}


\begin{thebibliography}{99}
\bibitem[\protect\citeauthoryear{Abbas \& Sheth}{2005}]{abbas}
 Abbas U., Sheth R.~K. 2005, MNRAS, 364, 1327
\bibitem[\protect\citeauthoryear{Allen, Evrard \& Mantz}{2011}]{allen}
  Allen S.W., Evrard A.E., Mantz A.B., 2011, ARA\&A, 49, 409
\bibitem[\protect\citeauthoryear{Allevato et al.}{2012}]{allevato}
  Allevato V. et al. 2012, ApJ, 758, 47
\bibitem[\protect\citeauthoryear{Ascasibar \& Diego}{2008}]{ascasibar}
 Ascasibar Y., Diego J.~M. 2008, MNRAS, 383, 369
\bibitem[\protect\citeauthoryear{Beers et al.}{1990}]{beers}
 Beers T.~C., Flynn K., Gebhardt K. 1990, AJ, 100, 32 
\bibitem[\protect\citeauthoryear{Behroozi et al.}{2015}]{behroozi} 
 Behroozi P.~S. et al. 2015, MNRAS, 454, 3020
\bibitem[\protect\citeauthoryear{Bianconi et al.}{2017}]{bianconi}
 Bianconi M. et al. 2017, MNRAS submitted 
\bibitem[\protect\citeauthoryear{B\"{o}hringer et al.}{2004}]{bohringer}
  B\"{o}hringer H. et al. 2004, A\&A, 425, 367
\bibitem[\protect\citeauthoryear{B\"{o}hringer et al.}{2014}]{bohringer14}
  B\"{o}hringer H., Chon G., Collins C.A. 2014, A\&A, 570, 31
\bibitem[\protect\citeauthoryear{Bonamente et al.}{2013}]{bonamente} 
 Bonamente M., Landry D., Maughan B., Giles P., Joy M., Nevalainen
 J. 2013, MNRAS, 428, 2812
\bibitem[\protect\citeauthoryear{Bond et al.}{1991}]{bond} 
 Bond J.R., Cole S., Efstathiou G., Kaiser N., 1991, ApJ, 379, 440
\bibitem[\protect\citeauthoryear{Bower}{1991}]{bower} 
 Bower R.G., 1991, MNRAS, 248, 332
\bibitem[\protect\citeauthoryear{Boylan-Kolchin et al.}{2009}]{boylan}
 Boylan-Kolchin M., Springel V., White S.D.M., Jenkins A., Lemson G.,
 2009, MNRAS, 398, 1150 
\bibitem[\protect\citeauthoryear{Correa et al.}{2015}]{correa}
 Correa C.~A., Wyithe S.~B., Schaye J., Duffy A.~R., 2015, MNRAS, 450, 1514
\bibitem[\protect\citeauthoryear{Chon et al.}{2012}]{chon12}
 Chon G., B\"{o}hringer H., Smith G.~P. 2012, A\&A, 548, 59
\bibitem[\protect\citeauthoryear{Chon et al.}{2013}]{chon13}
 Chon G., B\"{o}hringer H., Nowak N. 2013, MNRAS, 429, 3272
\bibitem[\protect\citeauthoryear{Chon \& B\"{o}hringer}{2015}]{chon}
 Chon G., B\"{o}hringer H., 2015, A\&A, 574, 132
\bibitem[\protect\citeauthoryear{Chung et al.}{2011}]{chung}
 Chung S.~M. et al. 2011, ApJ, 743, 34
\bibitem[\protect\citeauthoryear{Clerc et al.}{2016}]{clerc}
 Clerc N. et al. 2016, MNRAS, 463, 4490
\bibitem[\protect\citeauthoryear{Czoske}{2004}]{czoske}
 Czoske O. 2004, in IAU Colloq. 194, Outskirts of Galaxy Clusters:
 Intense Life in the Suburbs, ed. A. Diaferio (Cambridge: Cambridge
 Univ. Press), 183
\bibitem[\protect\citeauthoryear{David \& Kempner}{2004}]{david}
 David L.~P., Kempner J. 2004, ApJ, 613, 831
\bibitem[\protect\citeauthoryear{De Boni et al.}{2016}]{deboni}
 De Boni C., Serra A., Diaferio A., Giocoli C., Baldi M. 2016,
 ApJ, 818, 188
\bibitem[\protect\citeauthoryear{De Grandi et al.}{2016}]{degrandi}
 De Grandi S. et al. 2016, A\&A, 592, 154
\bibitem[\protect\citeauthoryear{Diaferio \& Geller}{1997}]{diaferio}
 Diafero A., Geller M.~J., 1997, ApJ, 481, 633
\bibitem[\protect\citeauthoryear{Dressler et al.}{2013}]{dressler}
 Dressler A., Oemler A.~Jr., Poggianti B., Gladders M.~D., Abramson
 L., Vulcani B. 2013, ApJ, 770, 62
\bibitem[\protect\citeauthoryear{Dutton \& Macci\`{o}}{2014}]{dutton}
  Dutton A.A., Macci\`{o} A.V., 2014, MNRAS, 441, 3359
\bibitem[\protect\citeauthoryear{Ebeling et al.}{1998}]{ebeling98}
 Ebeling H. et al. 1998, MNRAS, 301, 881
\bibitem[\protect\citeauthoryear{Ebeling et al.}{2000}]{ebeling00}
 Ebeling H. et al. 2000, MNRAS, 318, 333
\bibitem[\protect\citeauthoryear{Eckert et al.}{2014}]{eckert}
 Eckert D. et al. 2014, A\&A, 570, 119 
\bibitem[\protect\citeauthoryear{Eckert et al.}{2017}]{eckert17}
 Eckert D. et al. 2017, preprint (arXiv:1705.05844)
\bibitem[\protect\citeauthoryear{Edwards et al.}{2010}]{edwards}
 Edwards L.~O.~V., Fadda D., Frayer D.~T., Lima Neto G.~B., Durret
 F. 2010, ApJ, 140, 1891
\bibitem[\protect\citeauthoryear{Eisenstein \& Hu}{1998}]{eisenstein}
 Eisenstein D.~J., Hu W., 1998, ApJ, 496, 605 
\bibitem[\protect\citeauthoryear{Fadda et al.}{2008}]{fadda}
 Fadda D., Biviano A., Marleau F.~R., Storrie-Lombardi L.~J., Durret
 F. 2008, ApJL, 672, 9 
\bibitem[\protect\citeauthoryear{Fakhouri \& Ma}{2008}]{fakhouri08} 
 Fakhouri O., Ma C.-P., 2008, MNRAS, 386, 577
\bibitem[\protect\citeauthoryear{Fakhouri et al.}{2010}]{fakhouri10} 
 Fakhouri O., Ma C.-P., Boylan-Kolchin M., 2010, MNRAS, 406, 2267
\bibitem[\protect\citeauthoryear{Faltenbacher, Finoguenov \& Drory}{2010}]{faltenbacher}
 Faltenbacher A., Finoguenov A., Drory N. 2010, ApJ, 712, 484
\bibitem[\protect\citeauthoryear{Finoguenov et al.}{2007}]{finoguenov07}
 Finoguenov A. et al. 2007, ApJS, 172, 182
\bibitem[\protect\citeauthoryear{Finoguenov et al.}{2009}]{finoguenov09}
 Finoguenov A. et al. 2009, ApJ, 704, 564
\bibitem[\protect\citeauthoryear{Finoguenov et al.}{2010}]{finoguenov10}
 Finoguenov A. et al. 2010, MNRAS, 403, 2063
\bibitem[\protect\citeauthoryear{Finoguenov et al.}{2015}]{finoguenov15}
 Finoguenov A. et al. 2015, A\&A, 574, 130
\bibitem[\protect\citeauthoryear{Forman et al.}{1981}]{forman} 
 Forman W. et al. 1981, ApJL, 243, 133
\bibitem[\protect\citeauthoryear{Frenk et al.}{1999}]{frenk}
 Frenk C.~S. et al. 1999, ApJ, 525, 554 
\bibitem[\protect\citeauthoryear{Gao et al.}{2012}]{gao} 
 Gao L., Navarro J.F., Frenk C.S., Jenkins A., Springel V., White S.D.M., 2012, MNRAS, 425, 2169
\bibitem[\protect\citeauthoryear{Gehrels}{1986}]{gehrels}
 Gehrels N., 1986, ApJ, 303, 336 
\bibitem[\protect\citeauthoryear{Genel et al.}{2010}]{genel10} 
 Genel S., Bouch\'{e} N., Naab T., Sternberg A., Genzel R., 2010, ApJ, 719, 229
\bibitem[\protect\citeauthoryear{Giles et al.}{2012}]{giles}
 Giles P.~A., Maughan B.~J., Birkinshaw M., Worrall D.~M., Lancaster
 K., 2012, MNRAS, 419, 503 
\bibitem[\protect\citeauthoryear{Giocoli, Tormen \& van den Bosch}{Giocoli et al.}{2008}]{giocoli08} 
 Giocoli C., Tormen G., van den Bosch F.~C., 2008, MNRAS, 386, 2135
\bibitem[\protect\citeauthoryear{Giocoli et al.}{2010}]{giocoli10} 
 Giocoli C., Tormen G., Sheth R.~K., van den Bosch F.~C., 2010, MNRAS, 404, 502
\bibitem[\protect\citeauthoryear{Governato et al.}{1999}]{governato}
 Governato F., Babul A., Quinn T., Tozzi P., Baugh C.M., Katz N., Lake
 G. 1999, MNRAS, 307, 949
\bibitem[\protect\citeauthoryear{Gunn \& Gott}{1972}]{gunn}
 Gunn J.~E., Gott J.~R. 1972, ApJ, 176, 1
\bibitem[\protect\citeauthoryear{Gutierrez \& Krawczynski}{2005}]{gutierrez}
 Gutierrez K., Krawczynski H., 2005, ApJ, 619, 161
\bibitem[\protect\citeauthoryear{Haines et al.}{2007}]{haines07}
 Haines et al., 2007, MNRAS, 381, 7
\bibitem[\protect\citeauthoryear{Haines et al.}{2009a}]{haines09}
 Haines et al., 2009a, ApJ, 704, 126
\bibitem[\protect\citeauthoryear{Haines et al.}{2009b}]{haines09b}
 Haines et al., 2009b, MNRAS, 396, 1297
\bibitem[\protect\citeauthoryear{Haines et al.}{2010}]{haines10}
 Haines et al., 2010, A\&A, 518, L19
\bibitem[\protect\citeauthoryear{Haines et al.}{2012}]{haines12}
 Haines et al., 2012, ApJ, 754, 97
\bibitem[\protect\citeauthoryear{Haines et al.}{2013}]{haines13}
  Haines et al., 2013, ApJ, 775, 126
\bibitem[\protect\citeauthoryear{Haines et al.}{2015}]{haines15}
  Haines et al., 2015, ApJ, 806, 101
\bibitem[\protect\citeauthoryear{Harvey et al.}{2014}]{harvey}
 Harvey D. et al. 2014, MNRAS, 441, 404
\bibitem[\protect\citeauthoryear{Harvey et al.}{2015}]{harvey15}
 Harvey D. et al. 2015, Sci, 347, 1462
\bibitem[\protect\citeauthoryear{Henry et al.}{2009}]{henry}
 Henry J.~P., Evrard A.~E., Hoekstra H., Babul A., Mahdavi A. 2009,
 ApJ, 691, 1307
\bibitem[\protect\citeauthoryear{Hikage \& Yamamoto}{2016}]{hikage}
 Hikage C., Yamamoto K., 2016, MNRAS, 455, L77
\bibitem[\protect\citeauthoryear{Ichinohe et al.}{2015}]{ichinohe}
 Ichinohe Y. et al. 2015, MNRAS, 448, 2971
\bibitem[\protect\citeauthoryear{Jaff\'{e} et al.}{2016}]{jaffe}
 Jaff\'{e} Y.~L. et al. 2016, MNRAS, 461, 1202
\bibitem[\protect\citeauthoryear{Jiang \& van den Bosch}{2014}]{jiang}
 Jiang F., van den Bosch F.~C., 2014, MNRAS, 440, 193
\bibitem[\protect\citeauthoryear{Jones \& Forman}{1999}]{jones}
  Jones C., Forman W., 1999, ApJ, 511, 65
\bibitem[\protect\citeauthoryear{Just et al.}{2015}]{just}
 Just D.~W. et al. 2015, preprint (arXiv:1506.02051)
\bibitem[\protect\citeauthoryear{Kettula et al.}{2015}]{kettula}
 Kettula K. et al. 2015, MNRAS, 451, 1460
\bibitem[\protect\citeauthoryear{Kravtsov \& Borgani}{2012}]{kravtsov}
  Kravtsov A.V., Borgani S., 2012, ARA\&A, 50, 353
\bibitem[\protect\citeauthoryear{Lacey \& Cole}{1993}]{lacey} 
 Lacey C.~G., Cole S., 1993, MNRAS, 262, 627
\bibitem[\protect\citeauthoryear{Leauthaud et al.}{2010}]{leauthaud} 
Leauthaud A. et al., 2010, ApJ, 709, 97
\bibitem[\protect\citeauthoryear{Lemson \& Kauffmann}{1999}]{lemson}
 Lemson G., Kauffmann G., 1999, MNRAS, 302, 111  
\bibitem[\protect\citeauthoryear{Lemze et al.}{2013}]{lemze} 
 Lemze D. et al., 2013, ApJ, 776, 91
\bibitem[\protect\citeauthoryear{Lovisari, Reiprich \&
    Schellenberger}{Lovisari et al.}{2015}]{lovisari}
 Lovisari L., Reiprich T.~H., Schellenberger G., 2015, A\&A, 573, 118
\bibitem[\protect\citeauthoryear{Mahdavi et al.}{2013}]{mahdavi}
 Mahdavi A., Hoekstra H., Babul A., Bildfell C., Jeltema T., Henry
 J.~P. 2013, ApJ, 767, 116    
\bibitem[\protect\citeauthoryear{Mann \& Ebeling}{2012}]{mann}   
Mann A.~W., Ebeling H., 2012, MNRAS, 420, 2120
\bibitem[\protect\citeauthoryear{Mantz et al.}{2010}]{mantz10}   
Mantz A., Allen S.W., Rapetti D., Ebeling H., 2010, MNRAS, 406, 1759
\bibitem[\protect\citeauthoryear{Martinet et al.}{2016}]{martinet}
 Martinet N. et al. 2016, A\&A, 590, 69
\bibitem[\protect\citeauthoryear{Martino et al.}{2014}]{martino}
 Martino R. et al. 2014, MNRAS, 443, 2342 
\bibitem[\protect\citeauthoryear{McBride, Fakhouri \& Ma}{McBride et al.}{2009}]{mcbride}  
McBride J., Fakhouri O., Ma C.-P., 2009, MNRAS, 398, 1858
\bibitem[\protect\citeauthoryear{McCarthy et al.}{2008}]{mccarthy}
 McCarthy I.~G. et al. 2008, MNRAS, 383, 593 
\bibitem[\protect\citeauthoryear{McCarthy et al.}{2017}]{mccarthy17}
 McCarthy I.~G., Schaye J., Bird S., Le Brun A.~M.~C. 2017, MNRAS, 465, 2936
\bibitem[\protect\citeauthoryear{McGee et al.}{2009}]{mcgee} 
McGee S.~L., Balogh M.~L., Bower R.~G., Font A.~S., McCarthy
I.~G. 2009, MNRAS, 400, 937
\bibitem[\protect\citeauthoryear{Merloni et al.}{2012}]{merloni}
 Merloni A. et al. 2012, preprint (arXiv:1209.3314)
\bibitem[\protect\citeauthoryear{Mo \& White}{1996}]{mo}
 Mo H.~J., White S.~D.~M. 1996, MNRAS, 282, 347
\bibitem[\protect\citeauthoryear{Okabe \& Umetsu}{2008}]{okabe08}
 Okabe N., Umetsu K. 2008, PASJ, 60, 345   
\bibitem[\protect\citeauthoryear{Okabe et al.}{2010}]{okabe}   
 Okabe N., Takada M., Umetsu K., Futamase T., Smith G.P., 2010, PASJ, 62, 811
\bibitem[\protect\citeauthoryear{Okabe et al.}{2013}]{okabe13}
 Okabe N., Smith G.~P., Umetsu K., Takada M., Futamase T. 2013, ApJL, 769, 35
\bibitem[\protect\citeauthoryear{Okabe et al.}{2014}]{okabe14}
 Okabe N., Futamase T., Kajisawa M., Kuroshima R. 2014, ApJ, 784, 90
\bibitem[\protect\citeauthoryear{Okabe \& Smith}{Okabe et al.}{2016}]{okabe15}
 Okabe N., Smith G.~P., 2016, MNRAS, 461, 3794 
\bibitem[\protect\citeauthoryear{O'Sullivan et al.}{2017}]{osullivan}
 O'Sullivan E. et al. 2017, preprint (arXiv:1708.03555)
\bibitem[\protect\citeauthoryear{Pacaud et al.}{2016}]{pacaud}
 Pacaud F. et al. 2016, A\&A, 592, 2 
\bibitem[\protect\citeauthoryear{Pearson et al.}{2017}]{pearson}
 Pearson R. et al. 2017, MNRAS, 469, 3489 
\bibitem[\protect\citeauthoryear{Pedersen \& Dahle}{2007}]{pedersen}
 Pedersen K., Dahle H., 2007, ApJ, 667, 26  
\bibitem[\protect\citeauthoryear{Pereira et al.}{2010}]{pereira}
 Pereira M.~J. et al. 2010, A\&A, 518, L40 
\bibitem[\protect\citeauthoryear{Pierre et al.}{2016}]{pierre}
 Pierre M. et al. 2016, A\&A, 592, 1 
\bibitem[\protect\citeauthoryear{Planck Collaboration et al.}{2014a}]{planckXX}  
Planck Collaboration et al., 2014a, A\&A, 571, 20
\bibitem[\protect\citeauthoryear{Planck Collaboration et al.}{2014b}]{planckXVI} 
Planck Collaboration et al., 2014b, A\&A, 571, 16
\bibitem[\protect\citeauthoryear{Planck Collaboration et al.}{2016}]{planck2015} 
Planck Collaboration et al., 2016, A\&A, 594, 13
\bibitem[\protect\citeauthoryear{Poole}{2006}]{poole} 
 Poole G.~B., Fardal M.~A., Babul A., McCarthy I.~G., Quinn T.,
 Wadsley J. 2006, MNRAS, 373, 881
\bibitem[\protect\citeauthoryear{Press \& Schechter}{1974}]{press} 
 Press W.H., Schechter P., 1974, ApJ, 187, 425
\bibitem[\protect\citeauthoryear{Ragone et al.}{2004}]{ragone}
 Ragone C.~J., Merch\'{a}n M., Muriel H., Zandivarez A., 2004, MNRAS, 350, 983
\bibitem[\protect\citeauthoryear{Randall, Sarazin \& Ricker}{2002}]{randall}
 Randall S.~W., Sarazin C.~L., Ricker P.~M. 2002, ApJ, 577, 579
\bibitem[\protect\citeauthoryear{Reiprich et al.}{2013}]{reiprich}
 Reiprich T.~H. et al. 2013, Space Science Reviews, 177, 195
\bibitem[\protect\citeauthoryear{Ricker \& Sarazin}{2001}]{ricker}
 Ricker P., Sarazin C.~L. 2001, ApJ, 561, 621
\bibitem[\protect\citeauthoryear{Rines et al.}{2013}]{rines}
 Rines K., Geller M.~J., Diaferio A., Kurtz M.~J. 2013, ApJ, 767, 15
\bibitem[\protect\citeauthoryear{Rizza et al.}{1998}]{rizza}
 Rizza E., Burns J.~O., Ledlow M.~J., Owen F.~N., Voges W., Bliton
 M. 1998, MNRAS, 301, 328
\bibitem[\protect\citeauthoryear{Roediger et al.}{2015a}]{roediger}
 Roediger E. et al. 2015a, ApJ, 806, 103
\bibitem[\protect\citeauthoryear{Roediger et al.}{2015b}]{roediger2}
 Roediger E. et al. 2015b, ApJ, 806, 104
\bibitem[\protect\citeauthoryear{Sanderson \& Ponman}{2010}]{sanderson10}
 Sanderson A.~J.~R., Ponman T.~J. 2010, MNRAS, 402, 65
\bibitem[\protect\citeauthoryear{Schellenberger \& Reiprich}{2005}]{schellenberger}
 Schellenberger G., Reiprich T.~H., 2015, A\&A, 583, L2
\bibitem[\protect\citeauthoryear{Schuecker et al.}{2003}]{schuecker}
  Schuecker P., B\"{o}hringer H., Collins C.A., Guzzo L., 2003, A\&A, 398, 867 
\bibitem[\protect\citeauthoryear{Scoville et al.}{2007}]{scoville}
 Scoville N. et al. 2007, ApJS, 172, 1
\bibitem[\protect\citeauthoryear{Sheth \& Tormen}{1999}]{sheth}
 Sheth R.~K., Tormen G. 1999, MNRAS, 308, 119
\bibitem[\protect\citeauthoryear{Smith et al.}{2005}]{smith05}
 Smith G.~P. et al. 2005, MNRAS, 359, 417
\bibitem[\protect\citeauthoryear{Smith et al.}{2010}]{smith10}
 Smith G.~P. et al. 2010, A\&A, 518, L18
\bibitem[\protect\citeauthoryear{Smith et al.}{2016}]{smith16}
 Smith G.~P. et al. 2016, MNRAS, 456, L74
\bibitem[\protect\citeauthoryear{Spergel et al.}{2003}]{spergel}
 Spergel D.~N., et al. 2003, ApJS, 148, 175
\bibitem[\protect\citeauthoryear{Springel et al.}{2005}]{springel}
  Springel V. et al. 2005, Nature, 435, 629
\bibitem[\protect\citeauthoryear{Taylor \& Babul}{2005}]{taylor05}
 Taylor J.~E., Babul A. 2005, MNRAS, 364, 515
\bibitem[\protect\citeauthoryear{Tinker et al.}{2010}]{tinker10}
 Tinker J.~L. et al. 2010, ApJ, 724, 878 
\bibitem[\protect\citeauthoryear{van den Bosch, Tormen \& Giocoli}{van den Bosch et al.}{2005}]{van05}
  van den Bosch F.~C., Tormen G., Giocoli C., 2005, MNRAS, 359, 1029
\bibitem[\protect\citeauthoryear{van den Bosch et al.}{2014}]{van14}
  van den Bosch F.~C., Jiang F., Hearin A., Campbell D., Watson D., Padmanabhan N., 2014, MNRAS, 445, 1713
\bibitem[\protect\citeauthoryear{Vikhlinin et al.}{1998}]{vik98}
 Vikhlinin A., McNamara B.~R., Forman W., Jones C., Quintana H., Hornstrup A., 1998, ApJ, 502, 558
\bibitem[\protect\citeauthoryear{Vikhlinin et al.}{2009}]{vik09}
  Vikhlinin A. et al., 2009, ApJ, 692, 1060 
\bibitem[\protect\citeauthoryear{Voit}{2005}]{voit}
 Voit G.~M., 2005, Rev. Modern Phys., 77, 207 
\bibitem[\protect\citeauthoryear{Weinmann et al.}{2006}]{weinmann}
 Weinmann S.~M., van den Bosch F.~C., Yang X., Mo H.~J. 2006, MNRAS, 366, 2
\bibitem[\protect\citeauthoryear{Woo et al.}{2013}]{woo}
 Woo J. et al. 2013, MNRAS, 428, 3306
\bibitem[\protect\citeauthoryear{Xu et al.}{2015a}]{xu}
 Xu L., Rieke G.~H., Egami E., Pereira M.~J., Haines C.~P., Smith G.~P., 2015a, ApJS, 219, 18
\bibitem[\protect\citeauthoryear{Xu et al.}{2015b}]{xu2}
 Xu L., Rieke G.~H., Egami E., Haines C.~P., Pereira M.~J., Smith G.~P., 2015b, ApJ, 808, 159
\bibitem[\protect\citeauthoryear{Yang et al.}{2007}]{yang07}
 Yang X. et al. 2007, ApJ, 671, 153
\bibitem[\protect\citeauthoryear{Yang et al.}{2011}]{yang}
 Yang X., Mo H.~J., Zhang Y., van den Bosch F.~C., 2011, ApJ, 741, 13
\bibitem[\protect\citeauthoryear{Zabludoff et al.}{1996}]{zabludoff96}
 Zabludoff A.~I. et al. 1996, ApJ, 466, 104
\bibitem[\protect\citeauthoryear{Zabludoff et al.}{1998}]{zabludoff98}
 Zabludoff A.~I. \& Mulchaey J.~S. 1998, ApJ, 496, 39
\bibitem[\protect\citeauthoryear{Zentner \& Bullock}{2003}]{zentner}
 Zentner A.~R., Bullock J.~S., 2003, ApJ, 598, 49
\bibitem[\protect\citeauthoryear{Zhang et al.}{2007}]{zhang07}
 Zhang Y.-Y. et al. 2007, A\&A, 467, 437
\bibitem[\protect\citeauthoryear{Zhang et al.}{2011}]{zhang11}
 Zhang Y.-Y. et al. 2011, A\&A, 526, 105
\bibitem[\protect\citeauthoryear{Zhao et al.}{2009}]{zhao09}
 Zhao D.H., Jing Y.P., Mo H.J., B\"{o}rner G., 2009, ApJ, 707, 354
\bibitem[\protect\citeauthoryear{Ziparo et al.}{2014}]{ziparo}
 Ziparo F. et al. 2014, MNRAS, 437, 458
\end{thebibliography}
\end{document}